\documentclass{aa}  

\usepackage{graphicx}
\usepackage{txfonts}
\begin{document}
    \title{The K giant star Arcturus:\\ 
           the hybrid nature of its infrared spectrum
    \thanks{Tables 2 is only available in
            electronic form at the CDS via anonymous ftp to
            cdsarc.u-strasbg.fr (130.79.128.5) pub/A+A/vol/page or via
            http://cdsweb.u-strasbg.fr/cgi-bin/qcat?J/A+A/***/***.} 
    }

%    \subtitle{  }

   \author{T. Tsuji}

   \offprints{T. Tsuji}

   \institute{Institute of Astronomy, School of Sciences, 
          The University of Tokyo, Mitaka, Tokyo, 181-0015 Japan\\
         \email{ttsuji@ioa.s.u-tokyo.ac.jp}
    }

   \date{Received ; accepted }

% \abstract{}{}{}{}{} 
% 5 {} token are mandatory
 
  \abstract
   {}
   {We investigate the infrared spectrum of Arcturus 
   to clarify the nature of the cool component of its atmosphere,
   referred to as the CO-mosphere, and its relationship to the warm
   molecular envelope or the MOLsphere in cooler M (super)giant stars.  
   }
   {We apply the standard methods of spectral analysis 
    to the CO lines measured from the ``Infrared Atlas of the Arcturus 
    Spectrum'' by Hinkle, Wallace, and Livingston.
   } 
   {We found that the intermediate-strength lines (with  $-4.75 
   < {\rm log}\,{W/\nu} \la -4.4$: $W$ is the equivalent width and 
   $\nu$ the wavenumber) of CO fundamentals as well as overtones cannot be 
   interpreted with the line-by-line analysis based on the classical  line 
   formation theory, while the weaker lines can and provide some 
   information on the photosphere (e.g. log\,$A_{\rm C}$ = 
   7.97/log\,$A_{\rm H}$ = 12.00, $\xi_{\rm micro}$ = 
   1.87\,km\,s$^{-1}$, and $\xi_{\rm macro}$ = 3.47\,km\,s$^{-1}$).
   The nature of CO lines  shows an abrupt change at ${\rm log}\,{W/\nu} 
   \approx -4.75$ and the lines stronger than this limit indicate 
   large excess absorption that cannot be photospheric in origin. 
   This difficulty 
   also appears as an unpredictable upturn (at ${\rm log}\,{W/\nu} 
   \approx -4.75$) in the flat part of the curves-of-growth of CO lines.
   We confirm the same phenomenon in dozens of M giant stars, whose
   infrared spectra are known to have hybrid origins in 
   the photosphere and extra-molecular constituent referred to as 
   the MOLsphere. Thus the curve-of-growth analysis provides a simple means 
   by which to recognize the hybrid nature of the infrared spectra. 
   We note that the curves-of-growth of red supergiants and Mira
   variables found in the literature show similar peculiar patterns.
    The intermediate-strength lines of CO in Arcturus show only minor 
    expansion ($\la 0.6$\,km\,s$^{-1}$) relative to
    the weak lines and  only small line-asymmetry, but the strong lines 
    of the  CO fundamentals  exhibit higher turbulent velocity 
     than the other CO lines.    
    } 
   {The large excess absorption of the CO fundamental lines
    in Arcturus appears to be the same phenomenon as that found in the CO 
    overtone lines of cooler M giant stars.
    Thus, molecular condensation referred to as the MOLsphere in cool 
    luminous stars may also exist in Arcturus. The MOLsphere, however, 
    is not necessarily a separate ``sphere'' but may be an 
    aggregation of molecular clouds formed within the outer atmosphere. 
    The formation of molecular clouds (referred to as MOLsphere in our
    modeling) in the outer atmosphere appears to be a basic feature 
    of all the red giant stars from early K to late M types 
    (and red supergiants).
     }

   \keywords{Line: formation -- stars: individual: Arcturus -- stars: late-type
                 -- stars: atmospheres -- stars: chromosphere --
                 stars: mass-loss -- Infrared: stars
             }

  \maketitle

\section{Introduction}

The K giant star Arcturus ($\alpha$ Bootis: K1.5IIIp) is a prototype of 
cool giant stars because of its brightness and has been studied
extensively with a variety of methods. Spectroscopic analyses have
been facilitated greatly by the availability of high resolution 
spectral atlases; not only in the optical region (e.g. Griffin 1968) 
but also in the infrared region (Hinkle et al. 1995). 
Also, the other Arcturus atlases are reviewed by Hinkle et al.\,(1995).  

The optical spectrum has been assumed to originate   
in the photosphere and its detailed analyses based on photospheric
models have provided useful information on chemical abundances and 
photospheric structure among others. The infrared spectrum, however, 
appeared not to be  so simple: As soon as high 
resolution spectra in the 5 $\mu$m-region could  be observed by 
the Fourier Transform Spectroscopy (FTS) at Kitt Peak National
Observatory (KPNO), it was  noticed  by Heasley et al.\,(1978) 
that the lines of the CO fundamental lines could not be interpreted 
with the  photospheric model alone nor with the homogeneous model 
of the hot chromosphere based on the analysis of the CaII K line 
(Ayres \& Linsky 1975). The conclusion of this pioneering work about the 
CO fundamental lines was the possible presence of an inhomogeneous 
structure in the outer atmosphere of Arcturus (Heasley et al. 1978). 
  
After a detailed  analysis of the non-LTE effect on CO
line formation (Ayres \& Wiedemann 1989; Wiedemann \& Ayres 1991),
 further analysis of CO fundamental lines  by Wiedemann et al.\,(1994)
revealed that no direct chromospheric indicator was present in 
the CO fundamental lines. The CO lines instead exhibited a steady decrease
in temperature at the height where the chromospheric model shows
a temperature increase. Thus these authors proposed a thermal
bifurcation model for the outer atmosphere of Arcturus or
a possible presence of a cool ``CO-mosphere'' in addition to the hot 
chromosphere. Additional  evidence for a cool molecular 
constituent in Arcturus may be the detection of pure-rotation lines 
of H$_2$O in the mid-infrared region  by Ryde et al.\,(2002). 
Although these authors attributed the origin of the H$_2$O lines 
possibly to an anomalous structure of the photosphere, this 
marvelous discovery may also be a manifestation of the cool constituent 
in the outer atmosphere of Arcturus. We should  remember, however,
that an exact relationship between the water vapor lines and neither the 
anomalous photospheric structure nor the cool molecular constituent is 
known yet.    

The nature of the cool constituent, in either the  photosphere or 
the outer atmosphere, has not yet been understood even in Arcturus, 
which is a star relatively well studied. On the other hand, more or 
less similar phenomena were known, even more clearly, in cooler giant 
and supergiant stars. For example, it was shown that the infrared 
spectra of M giant stars represent a hybrid of at least
two components originating in the photosphere and a non-photospheric
molecular constituent (Tsuji 2008). This result  extended the 
idea of  quasi-static molecular layers suggested previously by the 
analysis of the low excitation CO lines alone (Tsuji 1988).
Also, H$_2$O lines were found in the early M giant (Tsuji 2001) and 
supergiant stars (Woolf et al. 1964; Tsuji 2000a), which should not be 
cool enough to accommodate water in their photospheres. These results again 
implied the possible presence of  an extra constituent referred to as a 
warm molecular envelope or a MOLsphere. Furthermore, direct evidence for
molecular layers in red (super)giant stars was shown by the 
extensive observations with spatial interferometry (e.g. Quirrenbach
et al. 1993; Perrin et al. 2004a, 2005, 2007).

We ask whether the inhomogeneous atmosphere or CO-mosphere
considered for Arcturus and the warm molecular envelope or MOLsphere
discussed for cooler M giant and supergiant stars are the same
phenomenon or not. There are two possible ways to answer this 
question: First we can investigate the 
nature of the infrared spectrum of Arcturus in some detail and look for
 similarities with the spectra of cooler stars. Second we can investigate
the CO fundamentals in M giant and supergiants to probe directly
the region of chromospheric activities and look for similarities with
the results for Arcturus.  The second approach would certainly
be attractive but would require formidable observational efforts if not
impossible with present-day observational facilities. In contrast, 
the first approach can be readily adopted thanks to the high quality
infrared spectral atlas of Arcturus by Hinkle et al.\,(1995), and we
follow this  approach in this paper. 

\section{Input data}
 We summarize the input data to be used in our analyses, namely
the spectra we use (Sect.\,2.1) and their measurements (Sect.\,2.2),
basic stellar parameters (Sect.\,2.3), and models we apply (Sect.\,2.4).  

\subsection{ Observed data}
 
We use the electronic version of the ``Infrared Atlas of the
Arcturus Spectrum'' by Hinkle et al.\,(1995). We analyze the spectra
ratioed to the telluric spectrum derived either from solar
or lunar spectra. Thanks to this difficult processing, as discussed 
in detail by Hinkle et al.\,(1995), many lines especially of the CO 
fundamentals could  be made available. The observations 
were repeated twice in general, one in winter and one in summer
(at KPNO in the northern hemisphere), to measure as 
many lines as possible in the spectra for different geocentric 
velocity shifts.      
    
We reproduce some detail of the spectra that we use in our Table 1 
which was adopted from Table 3 of Hinkle et al.\,(1995)  for easy 
reference. For the frequency bands 1867-1965  and 1965-2021\,cm$^{-1}$
 in Table 1, only  
winter data are available. For the frequency bands 2021-2098 
and 2119-2189\,cm$^{-1}$, both winter and summer data are 
available, and we treat them as 
independent data sets rather than  using the sum of the two. For 
the frequency band 4000-6675\,cm$^{-1}$,
we use  only summer data, although a winter spectrum is
also available, because a sufficient number of lines could  be
measured from the  summer spectrum alone. 

 \begin{table} 
\centering
\caption{  Observed data from the Arcturus IR Atlas (Hinkle et al.\,1995)}
\vspace{-2mm}
\begin{tabular}{  c c c c c }
\hline \hline
\noalign{\smallskip}
sp. ref.  &     freq. band  &    Res. &   obs. date  &    shift\\
    &     (cm$^{-1}$) &  (cm$^{-1}$) &        &  (km\,s$^{-1}$) \\
\noalign{\smallskip}
\hline
\noalign{\smallskip}
1   &   1867-1965   &    0.02 &   1994 Jan. 9 &  -30.52 \\
2   &   1965-2021  &   0.02  &  1994 Jan. 21 & -30.52   \\
3   &   2021-2098  &   0.02  &  1994 Jan. 7 &  -31.66   \\
4   &   2119-2189  &   0.02  &  1994 Jan. 7 &  -31.39 \\
\noalign{\smallskip}
5   &   2021-2098   &    0.02 &   1993 Jun. 7 &   14.63  \\
6   &   2119-2189   &   0.02  &  1993 Jun. 7  &  14.90  \\
7   &   4000-6675   &    0.04  &  1993 Jun. 5  &  14.33  \\
\noalign{\smallskip}
\hline
\end{tabular}
\vspace{5mm}
\end{table}

\begin{figure}
\centering
\includegraphics[width=8.5cm]{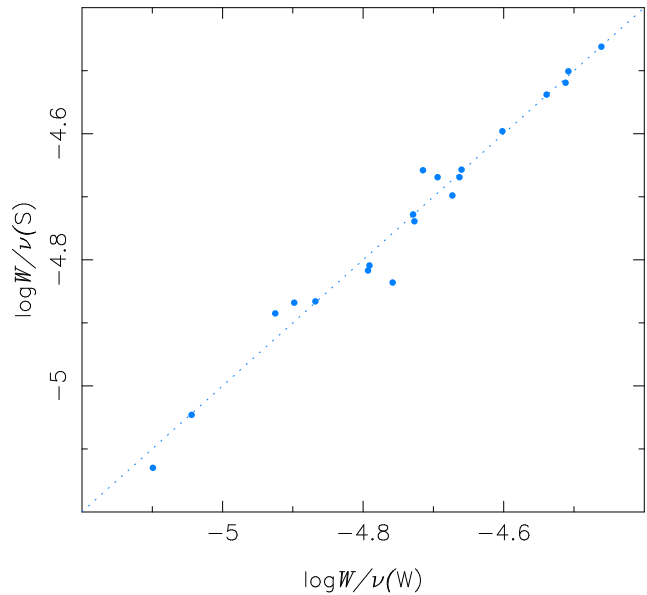}
\caption{
The values of log\,$W/\nu$ of the CO fundamental lines measured from the 
summer spectrum plotted against those from the winter spectrum.
}
\label{Fig1.eps}
\end{figure}

\subsection{ Measurements}
The continuum level of the ratioed spectra was adjusted to unity by
the authors of the atlas, but this may not imply that this level of
unity can be used as the continuum to which we refer in the measurements
of the line intensities. Many peaks
are found above the level of unity, especially in the $M$ band region
where disturbance by the atmospheric lines is most serious.     
We then detect the highest peak in every 2\,cm$^{-1}$ interval 
throughout the region of interest.  Since we  use the ratioed spectrum, 
it is possible that the detected  apparent  peaks are  disturbed  by
atmospheric lines, and we exclude such peaks. Thus, we plot the peaks
that are not seriously disturbed by the atmospheric absorption against 
wavenumber.  We fit a smooth curve passing these peaks but not
necessarily connecting all the peaks, since some peaks are apparently 
below the neighboring peaks. It is uncertain whether the  continuum 
adopted in this way is a true continuum, but we have no other reference 
by which  to measure equivalent widths. 

  With the fiducial ``continuum'' defined this way, we measure the line 
widths (FWHMs), depths, and equivalent widths (EWs) and the results are 
given in Table 2 (electronic form). We note that the instrumental effect
on the line widths and depths has not yet been corrected in Table 2. We 
measure the summer and winter spectra for the wavelength range of
2021-2098  and 2119-2189 cm$^{-1}$ (see  Table 1) separately and the 
resulting EWs measured in both spectra are compared in Fig.\,1, which
may provide some idea of the accuracy of our EW measurement.

%table 2 available electronically only
\onltab{2}{
\begin{table*}
\caption{Measured data of $^{12}$C$^{16}$O lines  in Arcturus Atlas (Only a few
 lines at the beginning of the table, as examples)}
\begin{tabular}{clccrcccrc}
 v'  v'' & rot. tr. & nu(cm-1) & log gf & L.E.P.(cm-1) & logEW/nu &
 depth & FWHM(cm-1) & log Gamma & sp. ref.(Tab.1) \\
--------&--------&----------&-------&---------------&---------&-------&
---------------&--------------&-------------\\
  2  1 & P   52 & 1875.683 & -2.962 & 7347.152  & -4.674 &  0.455 & 
0.082 & 0.751 & 1\\
  9  8 & P   14 & 1880.343 & -2.891 & 16782.602 & -5.025 &  0.265 &
 0.063 & -0.874 &1\\
 10  9 & P    7 & 1882.027 & -3.149 & 18440.730 & -5.128 &  0.200 &
 0.066 & -1.155 & 1\\
  3  2 & P   46 & 1882.874 & -2.839 &  8312.339 & -4.629 &  0.480 &
 0.087 & 0.584 & 1\\
  6  5 & P   30 & 1886.884 & -2.728 & 12153.510 & -4.833 &  0.378 &
 0.069 & -0.080 & 1\\
  9  8 & P   11 & 1892.263 & -2.994 & 16643.727 & -5.059 &  0.236 &
 0.066 & -0.847 & 1\\
  3  2 & P   43 & 1898.385 & -2.866 &  7809.389 & -4.641 &  0.486 &
 0.084 & 0.676 & 1\\
  9  8 & P    9 & 1900.043 & -3.080 & 16568.918 & -5.045 &  0.266 &
 0.061 & -0.832 & 1\\
  2  1 & P   45 & 1912.910 & -3.018 &  6060.500 & -4.575 &  0.515 &
 0.093 & 0.986 & 1\\
  2  1 & P   42 & 1928.441 & -3.045 &  5563.820 & -4.556 &  0.548 &
 0.092 & 1.078 & 1\\
\noalign{\smallskip}
\end{tabular}
\end{table*}
}

\subsection{ Basic stellar parameters}    
In Table 3, we summarize the basic stellar parameters of Arcturus used
as input data in our modeling and analyses. We mainly follow
a previous survey by Hinkle et al.\,(1995) summarized in their  Table 2, 
and update some data only if more recent results are available.
For example, the effective temperature of Arcturus appears to converge
to $\approx$4300\,K according to the recent result of angular diameter
measurement at $H$ band where the opacity reaches its minimum
(Lacour et al. 2008), and by the analysis of the spectral energy
distribution (Griffin \& Lynas-Gray 1999).  

\begin{table} 
\centering
\caption{ Basic stellar parameters of Arcturus }
\vspace{-2mm}
\begin{tabular}{  l l  }
\hline \hline
\noalign{\smallskip}
Parameter &    Value    \\
\noalign{\smallskip}
\hline
\noalign{\smallskip}
      angular diameter &  $ 21.05 \pm 0.21 $ (mas)$^{~a}$   \\
      effective temperature, $T_{\rm eff}$  & 4300 $\pm$ 30\,K$^{~b}$     \\
      surface gravity (log $g$) & $ 1.5 \pm  0.15$ $^{~b}$  \\
      ${\rm [Fe/H] }$  & -0.5 $\pm$ 0.1$^{~b}$       \\
      parallax &  $ 88.85 \pm 0.74 $ (mas)$^{~c}$      \\
      radius, $R_{*}$    &  25.5\,$ R_{\odot} \pm 1.5 $ $^{~d}$\\
      mass, $M_{*}$    &  $ 0.75\,M_{\odot} \pm 0.2^{~e} $    \\
      micro turbulent velocity, $\xi_{\rm micro}$ &1.85\,km\,s$^{-1}$ $^{~f}$\\
      radial velocity (heliocentric) &  -5.5\,km\,s$^{-1}$ $^{~g}$  \\
\hline
\noalign{\smallskip}
\end{tabular}
\begin{list}{}{}
\item[$^{\mathrm{a}}$]  Lacour et al. (2008); 
\item[$^{\mathrm{b}}$]  Peterson et al. (1993);
\item[$^{\mathrm{c}}$]  Hipparcos data (ESA 1997);
\item[$^{\mathrm{d}}$]  based on the angular diameter and parallax;
\item[$^{\mathrm{e}}$]  based on the radius and log\,$g$;
\item[$^{\mathrm{f}}$]  based on our preliminary analysis (our final
			value is 1.87\,km\,s$^{-1}$);
\item[$^{\mathrm{g}}$]  Griffin \& Griffin (1973).
\end{list}
\end{table}

\subsection{ Models of the photosphere and chromosphere }
    We  apply our model photosphere code,  used mainly for cooler red 
(super)giant stars (e.g. Tsuji 1978), to Arcturus  with the basic 
parameters discussed in Sect.\,2.3.  We assume the chemical abundances 
for 34 elements determined by Peterson et al. (1993). The 
resulting radiative equilibrium (RE) model starting integration from
log\,$\tau_{0} = -6.0$ ($\tau_{0}$ is the optical depth defined by the 
continuous opacity at 0.81\,$\mu$m) is shown by the solid line
in Fig.\,2 (model a).   Although our modeling code does not include 
atomic line-blanketing effect, our model agrees well with the model 
by Peterson et al. (1993) shown by the filled circles in Fig.\,2; the 
largest difference is about 50\,K at around $\tau_0 \approx 10^{-3}$.   
This result implies that the thermal structure of the photosphere
is largely determined by the blanketing effect of molecules rather
than of atoms even in K giants. We also compute a model assuming  
spherical symmetry (SS) rather than plane-parallel (PP). It is 
found that the differences between PP and SS models are rather minor 
for Arcturus photosphere, and a measure of the photospheric extension 
defined by $d = r(\tau_0 = 10^{-6})/R_{*}$ (where $R_{*}$ is the stellar 
radius) is only 1.04. For this reason, we apply PP models to our spectral 
analysis in this paper.  

We also compute an extended model photosphere (SS) by
starting integration from log\,$\tau_{0} = -9.0$ instead of
-6.0 adopted  in our modeling in general (e.g. model a).
We assume a turbulent velocity of 6\,km\,s$^{-1}$ in the
layers above log\,$\tau_{0} = -6.0$ and the resulting RE model
(model b in Fig.\,2) is in hydrostatic equilibrium with turbulent pressure 
included (see e.g. Tsuji 2006). We note that the surface
temperature is decreased below 2000\,K, although this has little effect on
observed properties such as spectra, colors, and angular diameters,  
since the matter density is very low in the layers above log\,$\tau_{0} 
= -6.0$.  We  discuss this model further in Sect.\,6.2.

\begin{figure}
\centering
\includegraphics[width=8.5cm]{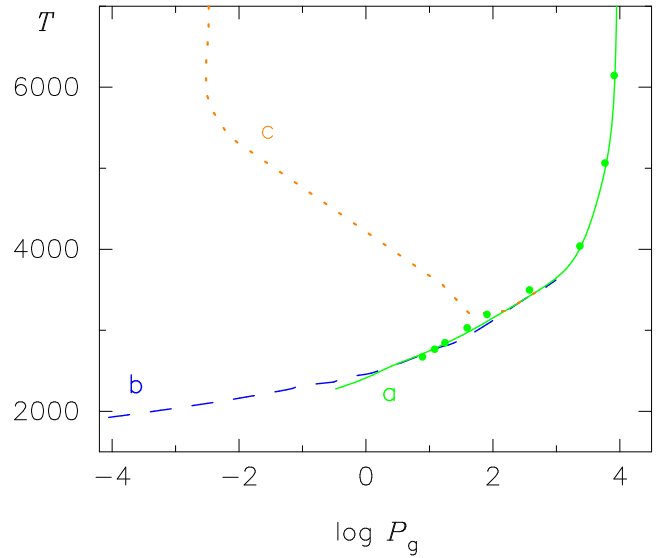}
\caption{
 Our RE model photosphere (plane parallel) of Arcturus
shown by the solid line (model {\bf a}) is compared with
the RE model by Peterson et al.(1993) shown by the filled circles. 
  Also,  an extended RE model (spherically symmetric) of 
Arcturus starting integration from log\,$\tau_{0} = -9.0$ rather 
than from log\,$\tau_{0} = -6.0$ (as in model {\bf a}) is shown 
by the dashed line (model {\bf b}). The chromospheric model by Ayres 
\&  Linsky (1975) is shown by the dotted line (model {\bf c}). 
}
\label{Fig2.eps}
\end{figure}
 
It is known that the photospheric model alone is insufficient
for interpreting the CO fundamental lines, which are formed
in the regime of the chromospheric activities. For this reason,
we fit the empirical model of the chromosphere of Arcturus
by Ayres \& Linsky (1975) to our photospheric PP model as shown 
in Fig.\,2 (model c).  

\section{ Line-by-line analysis of CO}
 We first apply a line-by-line (LL) analysis in which
each observed EW is interpreted with the use of the
classical microturbulent model of line formation.
For this purpose, we apply our CO linelist based on 
spectroscopic (Guelachivili et al. 1983) and intensity 
(Chackerian \& Tipping 1983) data.

\begin{figure}
\centering
\includegraphics[width=8.5cm]{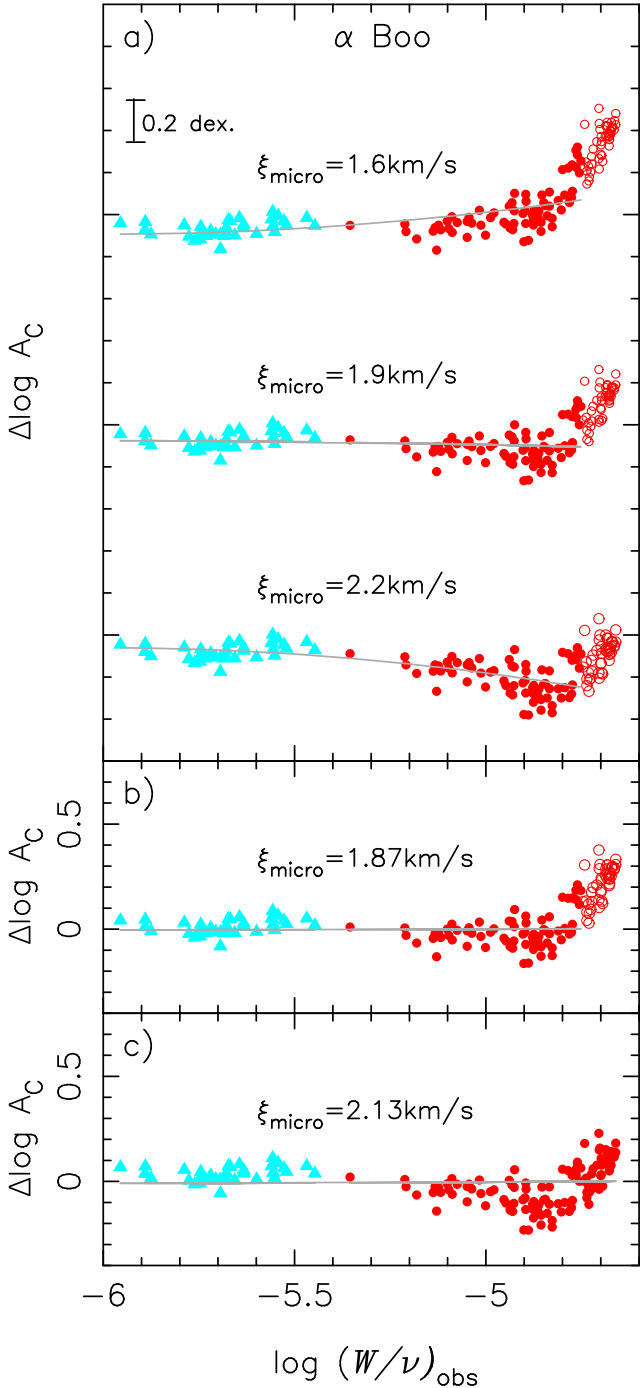}
\caption{
{\bf a)} 
Logarithmic abundance corrections  for the lines of the CO
 overtone bands observed in  $\alpha$ Boo plotted against
 $ {\rm log}\,(W/\nu)_{\rm obs} $ for assumed values of
$\xi_{\rm micro}$ = 1.6, 1.9, and 2.2 km\,s$^{-1}$. The CO 
first and second overtone lines are shown by the circles
and triangles, respectively. Note that the intermediate-strength lines 
(shown by the open symbols) are 
not included in the analysis for the reason detailed in the text.
(model photosphere: $T_{\rm eff}$/log\,$g$/$\xi_{\rm micro}$ =
4300/1.5/1.85).
{\bf b)}  
Confirmation of the null logarithmic  abundance corrections  for
log\,$A_{\rm C}$ = 7.97  and $\xi_{\rm micro}$ = 1.87 km\,s$^{-1}$, which
are the solution of the LL analysis of the CO weak lines (shown by the 
filled symbols in {\bf a}).
{\bf c)}  
Confirmation of the null logarithmic  abundance corrections  for
log\,$A_{\rm C}$ = 7.94  and $\xi_{\rm micro}$ = 2.13 km\,s$^{-1}$, which
are the formal solution of the LL analysis of all the CO overtone lines.
Note that the groups of lines with different strengths show inconsistent 
abundance corrections even though the mean logarithmic abundance 
correction is null.
}
\label{Fig3.eps}
\end{figure}

\subsection{The CO first and second overtone bands}
  We  apply a  detailed line-by-line analysis of the CO lines of
the first and second overtone bands. Given the model
photosphere as in Sect.\,2.4, a given equivalent width
can be interpreted in terms of the abundance and microturbulent
velocity. We start from the  abundances by Peterson et al. (1993).
In particular, the carbon abundance is ${\rm log}\,A_{\rm C} = 8.06$ 
on the scale of ${\rm log}\,A_{\rm H} = 12.00$ (we use this scale for
chemical abundances throughout this paper).  Then we assume a value 
of the microturbulent
velocity and determine an abundance correction to the assumed value 
needed to explain the observed equivalent width. This can be determined
from a relationship between  a few trial values of the abundance 
correction and the resulting equivalent widths (i.e. from a segment of 
a curve-of-growth, or a mini curve-of-growth, for the single line
we are to analyze). The resulting abundance
corrections should be the same for all the lines used  if a 
correct  microturbulent velocity is assumed. We 
repeat this process until we find such a value of the 
microturbulent velocity (as for detail, see Tsuji 1986). 

We show the resulting abundance corrections for the lines of the CO 
first (circles) and second (triangles) overtone bands obtained with the 
values of $\xi_{\rm micro}$ = 1.6, 1.9, and 2.2 km\,s$^{-1}$ in Fig.\,3a.
For most lines, the results follow as expected from the classical
line formation theory in that  the abundance corrections are
larger for smaller values of $\xi_{\rm micro}$ for saturated lines.
 We note, however, that the lines   of the CO
first overtone bands stronger than log\,$W/\nu \approx -4.75$ 
do not follow the expected behavior in that they never show  
abundance corrections consistent with those for the weaker lines 
for any assumed value of $\xi_{\rm micro}$. 

This is the same phenomenon that we have identified in many M giant stars,
and we concluded that these lines, which we referred to as the
intermediate-strength lines
\footnote{We referred to the lines of
log\,$W/\nu \la -4.75$ as the weak lines (but not necessarily so
weak as to be free from saturation effect), those with  $-4.75 < 
{\rm log}\,W/\nu \la -4.4$ as the intermediate-strength lines,
and those with log\,$W/\nu > -4.4$ as the strong lines (Tsuji 2008), 
and we follow this classification in this paper.}, 
should be contaminated with contributions of non-photospheric origin 
(Tsuji 2008). We now confirm exactly the same phenomenon in the K 
giant star Arcturus. For this reason, we disregard the lines of 
log\,$W/\nu > -4.75$ (i.e. the intermediate-strength lines) in our 
LL analysis, and obtain an abundance correction of $\Delta\,{\rm log}\,A = 
-0.09 \pm 0.01$ and $\xi_{\rm micro} = 1.87 \pm 0.02$\,km\,s$^{-1}$ from 
the lines with log\,$W/\nu \la -4.75$ (i.e. the weak lines). Since the 
CO abundance directly reflects the  carbon abundance in oxygen-rich 
giants, we derive a revised value of  ${\rm log}\,A_{\rm C} = 8.06 - 
0.09 = 7.97 $. We confirm in Fig.\,3b that this carbon abundance and 
$\xi_{\rm micro} = 1.87$\,km\,s$^{-1}$ provide a mean null logarithmic 
abundance correction. 

If we include all the measured lines in our LL analysis, 
it is also  possible to obtain formal solutions, which are 
 $\Delta\,{\rm log}\,A = -0.12 \pm 0.01$ 
and  $\xi_{\rm micro} = 2.13 \pm 0.02$\,km\,s$^{-1}$. The resulting
carbon abundance of ${\rm log}\,A_{\rm C} = 8.06 - 0.12 = 7.94$ and 
$\xi_{\rm micro} = 2.13$\,km\,s$^{-1}$ provide a mean null logarithmic
abundance correction as shown in Fig.\,3c,  but the different
groups of lines, i.e., those of the CO second overtones, the weak lines
of the CO first overtones, and the intermediate-strength lines   
of the CO first overtones, show large deviations from the null
correction. This result confirms that it is not possible to
carry out a consistent abundance analysis if we include the 
intermediate-strength lines in our LL analysis.

\subsection{The CO fundamental bands}
  We now include in our LL analysis the lines of the CO fundamentals 
shown by the squares in Fig.\,4a. It appears that 
the stronger and weaker lines, divided at about log\,$W/\nu \approx -4.75$, 
behave quite differently in the fundamentals more clearly than in the 
overtones. But the nature of the peculiar behavior of the stronger lines or
the intermediate-strength lines is essentially the same as those lines
of the overtones bands in Arcturus (see Fig.\,3) as well as in many M
giant  stars (Tsuji 2008). 

  We again disregard the lines with log\,$W/\nu > -4.75$ 
 in the fundamentals as well as in the overtones
in our LL analysis, and find $\Delta\,{\rm log}\,A = -0.11
 \pm 0.01$ and $\xi_{\rm micro} = 2.01 \pm 0.03$\,km\,s$^{-1}$.
The revised carbon abundance of ${\rm log}\,A_{\rm C} = 8.06 - 0.11 =
 7.95 $ together with $\xi_{\rm micro} = 2.01 $\,km\,s$^{-1}$  
result in a null mean logarithmic abundance
as confirmed in Fig.\,4b. Thus, by including the fundamental 
lines with log\,$W/\nu \la -4.75$ in our LL analysis, $\Delta\,{\rm
 log}\,A$ and $\xi_{\rm micro}$ change  by  5\% (from -0.09 to 
- 0.11 dex.) and 7\% (from 1.87 to 2.01\,km\,s$^{-1}$), respectively.
 For the origin of the differences caused by the inclusion
of the fundamental bands, we already know that the effect of the hot 
chromosphere should be considered in the analysis of the CO fundamentals 
in Arcturus (Heasely et al. 1978; Wiedemann \& Ayres 1994).
For this reason (and also for another reason to be discussed in
Sect.\,6.2), we propose that the results including the
fundamental lines do not represent the photospheric values, 
and we adopt the results based on the weak lines of the overtone bands
(Sect.\,3.1) as the photospheric abundance (i.e. ${\rm log}\,A_{\rm C} =
7.97 \pm 0.01$) and microturbulent velocity (i.e. $\xi_{\rm micro} = 
1.87 \pm 0.02$\,km\,s$^{-1}$) of Arcturus throughout this paper.  

\begin{figure}
\centering
\includegraphics[width=8.5cm]{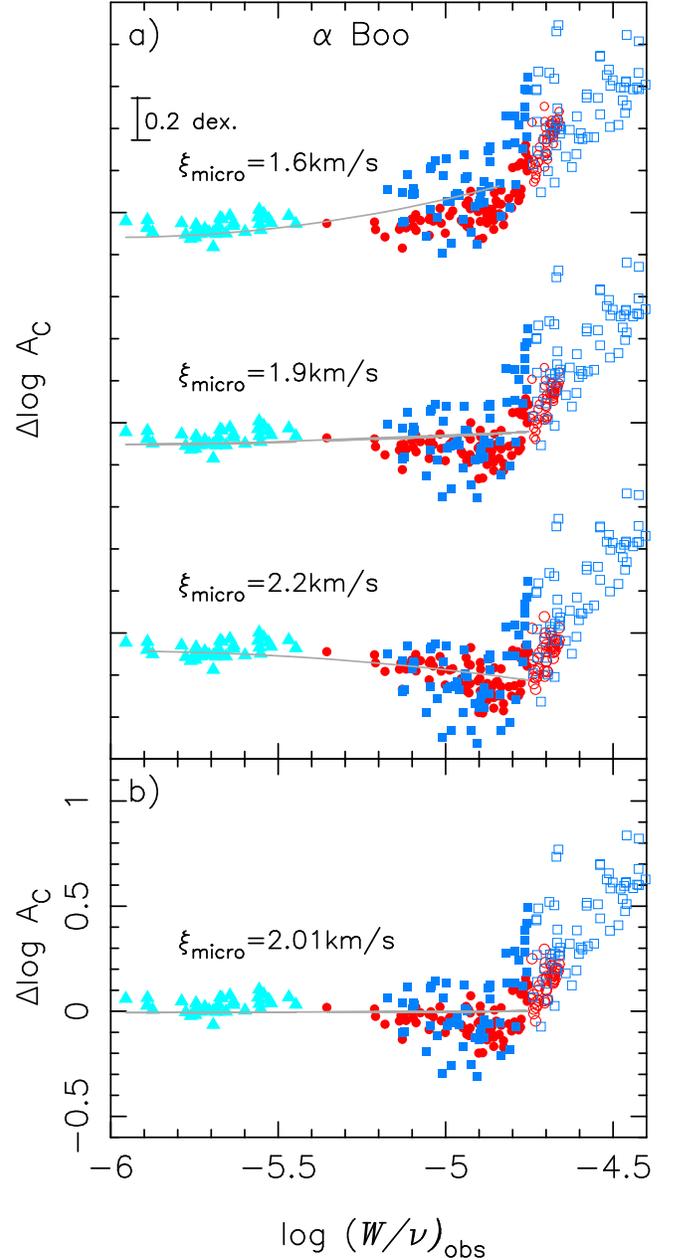}
\caption{
{\bf a)} The same as in Fig.\,3a, but  the lines of the CO fundamental bands
shown by the squares are added to the LL analysis.  
Note that the intermediate-strength lines (shown by the open symbols) are 
again not included in the analysis.
{\bf b)}  Confirmation of the null logarithmic  abundance corrections  for
log\,$A_{\rm C}$ = 7.95  and $\xi_{\rm micro}$ = 2.01 km\,s$^{-1}$, which
are the solution of the LL analysis of the weak lines of CO
(shown by the filled symbols in {\bf a}).
}
\label{Fig4.eps}
\end{figure}

The behavior of the lines stronger than about log\,$W/\nu \approx -4.75$
cannot be explained at all by the effect of the hot chromosphere. 
It appears that the excess absorption, possibly of non-photosphere
origin, is more prominent in the stronger lines of the fundamentals
than in the overtones. The detailed LL analysis outlined in this section, 
however, is not well optimized to the inhomogeneous models with cool 
and hot components, and we examine these cases with a different method 
in the next section. 

\section{ Curve-of-growth analysis}
It appears that the CO lines may not originate in the photosphere 
alone but may be disturbed not only by the effect of the hot chromosphere
but also by other cool constituents as in M giant stars (Tsuji 2008). In 
these circumstances, we return to a simpler method based on the 
curve-of-growth by which all the observed equivalent widths are 
interpreted in terms of a single curve-of-growth  for a set of 
representative parameters such as LEP (lower excitation potential), 
line position, and  damping constant. In contrast, we
essentially use different mini curves-of-growth for each individual
lines with specified LEP, wavenumber, and damping constant in our detailed
LL analysis
\footnote{We do not in fact consider the damping constants
in detail but apply a simple analytical formula for collision half-width
(Tsuji 1986), since collision damping plays a minor role in giant
stars.}. 
Thus, the accuracy may be lower in the curve-of-growth (CG) 
analysis, but we hope to obtain a more intuitive view by adopting a 
simpler method.      

\subsection{Arcturus}
   In Fig.\,5a, we plot the observed values of log\,$W/\nu$ (from 
Table 2) of the CO first (filled circles) and  
second (filled triangles) overtone lines  against
      $$  {\rm log}\,(W/\nu)_{\rm wk} =  {\rm log}\,gf
                        + {\rm log}\,\Gamma_{\nu}(\chi), \eqno(1) $$   
where $\Gamma_{\nu}(\chi)$ is the line intensity integral evaluated by 
the weighting function method (for details, see the Appendix of Tsuji 
1991). To this purpose, we apply our PP model photosphere (model a in 
Fig.\,2) with the carbon abundance of log\,$A_{\rm C} = 7.97$ determined 
in Sect.\,3.1. For comparison, we compute theoretical curves-of-growth 
for fictitious lines of LEP = 1 eV and log\,$gf = -7.00 \sim -2.00$ (by 
a step of 0.50) at $\lambda = 2.30\,\mu$m ($\nu$ = 4348 cm$^{-1}$) for 
the CO first overtones (solid line), and at $\lambda = 1.67\,\mu$m ($\nu$ = 
6000\,cm$^{-1}$) for the CO second overtones (dashed line)
\footnote
{The theoretical curve-of-growth is defined as a relationship
of ${\rm log}\,W/\nu$ against ${\rm log}\,(W/\nu)_{\rm wk}$  and
converges to ${\rm log}\,W/\nu = {\rm log}\,(W/\nu)_{\rm wk}$ at the
weak-line limit. The observed curve-of-growth is a plot of 
${\rm log}\,(W/\nu)_{\rm obs}$ against ${\rm log}\,(W/\nu)_{\rm wk}$,
and converges to  ${\rm log}\,(W/\nu)_{\rm obs}  
= {\rm log}\,(W/\nu)_{\rm wk}$ at the weak line limit only if the
abundances assumed in evaluating $ {\rm log}\,(W/\nu)_{\rm wk}$ are 
correct.}. 
In the theoretical curve-of-growth, we assume the microturbulent velocity 
of $\xi_{\rm micro}$ = 1.87\,km\,s$^{-1}$ determined in Sect.\,3.1.     
The curves-of-growth for different lines should differ slightly depending
on LEP and line position, and these differences
are considered in our LL analysis. We recall, however, that
these differences are ignored and one curve for representative
parameters is used for all the lines in the classical
CG method (coarse analysis). 

\begin{figure}
\centering
\includegraphics[width=8.5cm]{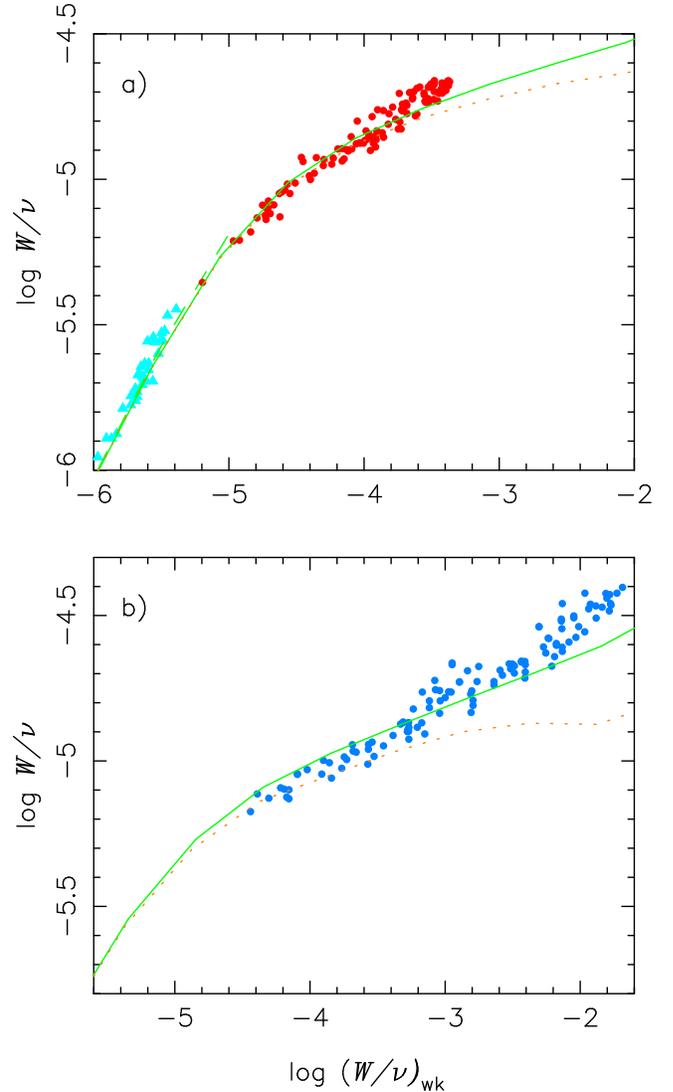}
\caption{
{\bf a)} Empirical curves-of-growth of the CO first (filled circles) and 
second (filled triangle) overtones observed in Arcturus are compared with the
theoretical ones (solid and dashed lines for the first and second
 overtones, respectively) based on the PP model and the microturbulent 
velocity $\xi_{\rm micro}$ = 1.87\,km\,s$^{-1}$. The result for the 
homogeneous chromosphere shown in Fig.\,2 is given by the dotted line.  
{\bf b)} Empirical curve-of-growth of the CO fundamentals (filled 
circles) is compared with the theoretical one (solid line) based on 
the PP model and the microturbulent velocity $\xi_{\rm micro}$ = 
1.87\,km\,s$^{-1}$. The result for the homogeneous chromosphere is 
shown by the dotted line.
}
\label{Fig5.eps}
\end{figure}

    Inspection of Fig.\,5a reveals that the observed and theoretical
curves-of-growth of the CO second overtones agree  well in the 
weak line limit, and this fact confirms that the carbon abundance 
determined in Sect.\,3.1 is supported by the most simple analysis 
of weak lines independently of the microturbulent model of line
formation.
The observed and theoretical curves-of-growth should also agree in the
flat part of the curves-of-growth, if the microturbulent velocity
assumed is correct. This expectation is fulfilled  only partly in the
region of $ {\rm log}\,W/\nu \la -4.75$, namely at about the same region
where our LL analysis could have been done consistently (Sect.3.1).
However, the empirical curve-of-growth deviates from the theoretical one
in the region of $ {\rm log}\,W/\nu > -4.75$ 
\footnote{In the classical CG analysis, the microturbulent velocity is 
determined  by searching a best fit  among  theoretical 
curves-of-growth of different microturbulent velocities. Then it may be 
possible that the empirical curve-of-growth such as Fig.\,5a may be 
fitted   with a theoretical curve-of-growth of the microturbulent 
velocity somewhat larger than 1.87\,km\,s$^{-1}$. With the result of the 
detailed line-by-line analysis in Sect.\,3.1 at hand, we could avoid such an 
erroneous fit and we already know that the peculiar behavior of the
intermediate-strength lines cannot be a problem of the value of the 
microturbulent velocity as noted in Sect.\,3.1 (also see Tsuji 2008). 
This may be a reason why the LL analysis should be preferred to the 
much simpler classical CG analysis.}. 
This result should be what can be expected from our LL analysis 
(Sect.\,3.1), and the peculiar behavior of the intermediate-strength 
lines is now confirmed  by a simple curve-of-growth analysis.   

 In Fig.\,5b, we compare  the empirical curve-of-growth for the CO 
fundamental lines (filled circles) with the theoretical one computed 
for fictitious lines of  LEP = 1 eV and log\,$gf = -7.00 \sim -2.00$
at $\lambda = 5.0\,\mu$m ($\nu$ = 2000\,cm$^{-1}$) (solid line). We 
again use our PP model photosphere with log\,$A_{\rm C} = 7.97$ and 
$\xi_{\rm micro}$ = 1.87\,km\,s$^{-1}$ determined in Sect.\,3.1. The 
observed  curve-of-growth is slightly below the theoretical one for 
lines  weaker than about $ {\rm log}\,W/\nu  \approx -4.85$ (about 
0.1\,dex lower than the critical value of -4.75) but shows opposite 
behavior for the stronger lines, consistent with the result of the 
LL analysis in Sect.\,3.2. 
 
So far, we have considered only the classical photosphere, but it has
been known that the hot chromosphere should have some effects especially
on the CO fundamentals formed in the upper photosphere (Heasely et al. 
1978; Wiedemann \& Ayres 1994). We compute theoretical 
curves-of-growth for the empirical model of the hot homogeneous 
chromosphere (model c in Fig.\,2) derived from the analysis of Ca II  H \& K 
emission (Ayres \& Linsky 1975) discussed in Sect.\,2.4, and the results 
are shown by the dotted lines in Figs.\,5a \& b. 
As expected, the lines are weakened by a temperature reversal 
in the chromospheric model (Fig.\,2). The effect, however, is
quite minor in the lines of the overtones, which are formed mostly
below the temperature minimum (Fig.\,5a). Departure from LTE 
in CO is also not important in the layers below the temperature minimum
in Arcturus (Ayres \& Wiedemann 1989).

    The effect of the hot chromosphere is increasingly prominent in the 
lines of the CO fundamentals (Fig.\,5b). The observed curve-of-growth
of the weaker lines appears between the theoretical curve-of-growth
based on the classical RE model photosphere (solid line) and that based 
on the model of the homogeneous chromosphere (dotted line). This result 
may imply  that the weaker lines of the CO fundamentals may be affected 
by the chromosphere, but only partly. This result is essentially the same
as the previous result  by Heaseley et al. (1978), in which the line
profiles of the weaker lines of the CO fundamentals could not be 
understood by any single component model, neither homogeneous 
chromosphere nor a classical photosphere, but could be better 
described by the theoretical spectrum from a two-component model
consisting of the hot chromosphere and cool photosphere. 

The difficulty in interpreting the stronger lines of the CO fundamentals
was discussed  by Wiedemann \& Ayres (1992) and by Wiedemann et al. 
(1994), who showed that the strong lines of the CO fundamentals indicate 
very low surface temperature near 2400\,K with no indication of the 
temperature increase expected by an empirical chromospheric model
such as that of Ayres \& Linsky (1975).
Based on a larger sample of CO lines, we arrive at a similar result   
in Fig.\,5b:  It is as if the  possible weakening of the stronger lines 
by the chromospheric temperature increase is superseded by another effect 
producing very strong absorption. Thus, we agree with the conclusion of 
Wiedemann and his collaborators that the outer atmosphere of Arcturus 
is dominated by the cool gaseous component, which was referred to as 
the  CO-mosphere by them. In view of this result that the effect of
the  chromosphere may not be significant, we  also consider  another
interpretation of the weaker CO fundamental lines in Sect.\,6.2, 
which we consider to be more likely. 

\begin{figure}
\centering
\includegraphics[width=8.5cm]{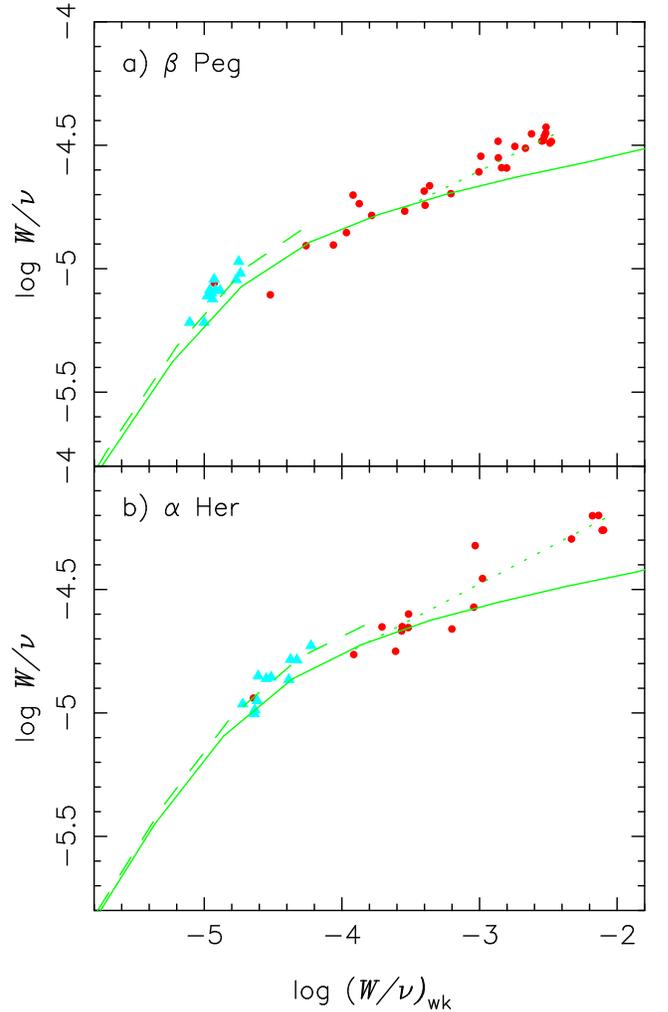}
\caption{
 Empirical curves-of-growth of the CO first (filled circles) and second 
(filled triangles) overtones  are compared with the
theoretical ones (solid and dashed lines are for the first and second
 overtones, respectively) based on the PP models and the microturbulent 
velocities  given in Table 4. Note that the lines of the
 intermediate-strength show larger equivalent widths than those
 predicted by the theoretical curves-of-growth.
{\bf a)} $\beta$ Peg (M2.5II-III). {\bf b)} $\alpha$ Her (M5Ib-II).
}
\label{Fig6.eps}
\end{figure}

\begin{figure}
\centering
\includegraphics[width=8.5cm]{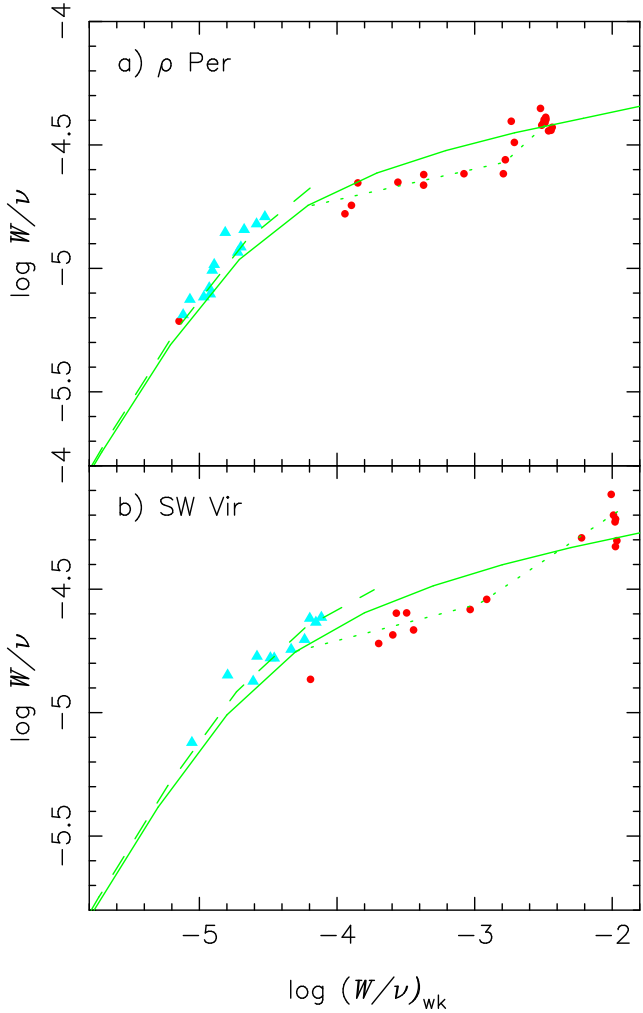}
\caption{
The same as Fig.\,6 but note that the 
 intermediate-strength lines show equivalent widths smaller than those
 predicted by the theoretical curves-of-growth.
{\bf a)}  $\rho$ Per (M4II).  {\bf b)} SW Vir (M7III:).  
}
\label{Fig7.eps}
\end{figure}
 
So far we have assumed LTE in our analysis throughout, but we referred
to the detailed analysis of the departure from LTE in CO line formation
by Ayres \& Wiedemann (1989). They showed that the non-LTE effect in CO
line formation is certainly non-negligible and even quite appreciable
in the surface layers. 
However, their detailed numerical simulation of the CO lines of the 
fundamental bands showed that the non-LTE absorption cores of strong 
CO lines are slightly deeper than the LTE spectrum based
on the RE model of Arcturus but are far too short to
explain the observed strong CO lines. They also showed that non-LTE and
LTE spectra for the chromospheric model (Ayres \& Linsky 1975) are
almost identical
\footnote{This unexpected result can be because
CO lines are formed near the temperature minimum where departure from
LTE is still minor and CO dissociates rapidly in the hot upper layers 
where non-LTE effect should be more significant. }. 
Thus, our conclusion based on an LTE analysis may be little affected by 
including a non-LTE effect, at least qualitatively.

\begin{table*} 
\label{table:1}
\caption{ Input data for the curves-of-growth of M giants based on PP
 model (results based on SS models in parenthesis) }
\vspace{-2mm}
\centering
\begin{tabular}{l c l c c c l l}
\hline \hline
\noalign{\smallskip}
object & molecule & PP model & $\xi_{\rm micro}^{\rm PP}$ 
($\xi_{\rm micro}^{\rm SS}$) & log\,$A_{\rm C}^{\rm PP}$ 
(log\,$A_{\rm C}^{\rm SS}$) & log\,$A_{\rm O}^{\rm PP}$ 
(log\,$A_{\rm O}^{\rm SS}$) &  CG analysis & LL analysis  \\
       &   &($T_{\rm eff}$/log\,$g$)  &  (km\,s$^{-1}$) &
 &   & (this Paper) & (Tsuji 2008)    \\
\noalign{\smallskip}
\hline
\noalign{\smallskip}
$\beta$ Peg & CO & 3600/0.5 & 1.88 $\pm$ 0.07 (2.04) & 
8.17 $\pm$ 0.05 (8.27)& - &  Fig\,.6a & Fig.\,6a,b \\
$\rho$ Per  & CO  &  3500/0.5 & 3.52 $\pm$ 0.73 (3.59) & 8.17 $\pm$ 0.08
(8.27) & -  & Fig.\,7a & Fig.\,8a,b \\
$\alpha$ Her  & CO & 3300/0.5 & 2.67 $\pm$ 0.71 (2.82) & 
8.46 $\pm$ 0.16 (8.40) & - & Fig.\,6b & Fig.\,4a,b \\
SW Vir  & CO & 2900/0.0 & 4.10 $\pm$ 3.00 (4.21) & 
8.22 $\pm$ 0.22 (8.26) & - &  Fig.\,7b & Fig.\,9c \\ 
\noalign{\smallskip}
\noalign{\smallskip}
$\alpha$ Her  & OH & 3300/0.5 & 2.57 $\pm$ 0.09 (2.92) & - & 
8.82 $\pm$ 0.05 (8.85) &  Fig.\,8a & Fig.\,10a,b  \\
SW Vir   &   OH  &  2900/0.0  & 1.69 $\pm$ 0.15 (1.83)  & - & 
8.42 $\pm$ 0.12 (8.54) & Fig.\,8b & Fig.\,11c\\ 
\noalign{\smallskip}
\hline
\noalign{\smallskip}
\end{tabular}
\end{table*}

\subsection{M-giant stars}
 The unpredictable upturn in the curve-of-growth of the CO overtones 
(Fig.\,5a) is rather subtle in the case of Arcturus, but the upturn may 
be more prominent in cooler M giant stars in which our LL
analysis revealed large anomalies in lines stronger than   
$ {\rm log}\,W/\nu \approx -4.75$ (Tsuji 2008). We now apply
the CG analysis to M giants, since this should provide a
simple means by which to investigate the nature of the
excess absorption (or emission) in cool giant stars.  

 For this purpose, we apply the CG analysis to a few M giants by 
using PP models as  in Arcturus instead of SS models as in Tsuji 
(2008), for simplicity. We also prefer PP models rather than
SS models because ${\rm log}\,(W/\nu)_{\rm wk}$ can be evaluated easily
by the weighting function method for PP models (e.g. Uns\"old 1955) but
no  such simple formula is known for SS models.
A few cases we analyzed as examples are summarized in Table 4
\footnote{Along with the results based on PP models, those based on
SS models are reproduced from Table 6 (Tsuji 2008) in Table 4 (in
parenthesis). In general, the differences between PP and SS analyses do not
differ significantly from the probable errors (almost the same for PP and
SS analyses and not shown for SS results) of the resulting abundances
and turbulent velocities.}. 
We proceed for each M giant as for Arcturus leading to
Fig.\,5a: We first plot observed  ${\rm log}\,W/\nu$ based
on the data in Table 3 (Tsuji 2008) against  ${\rm log}\,(W/\nu)_{\rm wk}$ 
evaluated with the  carbon abundance of Table 4 based on the PP model (hence
differs slightly from log\,$A_{\rm C}$ in Tsuji 2008).
The resulting empirical curve-of-growth is then compared with the 
theoretical one with $\xi_{\rm micro}$ of Table 4 based on the PP model.

The results for CO lines in $\beta$ Peg and $\alpha$ Her are
shown in Figs.\,6a \& 6b, respectively. The observed  
curves-of-growth of the first (filled circles) and second (filled
triangles) overtones  agree reasonably well with the theoretical ones
for the first (solid lines) and second (dashed lines) overtones,
respectively, for lines of ${\rm log}\,W/\nu \la -4.75$. However, the 
observed data begin to deviate from the trend expected for the flat 
part of the curves-of-growth at  ${\rm log}\,W/\nu \approx -4.75$ 
both in $\beta$ Peg and $\alpha$ Her. These deviations of the
intermediate-strength lines are quite consistent with our LL analysis of 
these objects (referred to in Col.\,8 of Table 4) in that these lines 
provided unreasonably large abundance corrections, which we have 
interpreted as being caused by excess absorption originating in 
the MOLsphere (Tsuji 2008).

The cases of $\rho$ Per and SW Vir are shown in Figs.\,7a \& 7b,
respectively. In these cases, the intermediate-strength lines also
show deviations from the expected  theoretical curves-of-growth
but in an opposite direction than for $\beta$ Peg and $\alpha$ Her
(Figs.\,6a \& 6b). These results, however, are again consistent
with our LL analysis of these objects (referred to in Col.\,8 of 
Table 4), which provided much smaller abundance corrections for the 
intermediate-strength lines than for the weak lines, and we have 
interpreted these results as due to weakening of the 
intermediate-strength lines by the emission of the MOLsphere (Tsuji 2008).

Finally, we examine the case of the OH fundamental lines observed in
the $L$ band region of M giant stars. We create curves-of-growth 
for $\alpha$ Her and SW Vir based on the EWs given in Table 3 (Tsuji
2008) and the results are shown in Figs.\,8a \& 8b, respectively. 
The observed and theoretical curves-of-growth are shown by the filled 
circles and solid lines, respectively.  The unpredictable upturn of the
observed curves-of-growth at  ${\rm log}\,W/\nu \approx -4.75$  appears 
more clearly for OH than for CO discussed 
above. This is simply because OH data are superior to CO data both in 
quality and quantity. Thus, the peculiar behaviors of the 
intermediate-strength lines noted by our LL analysis (referred to in 
Col.\,8 of Table 4) can be shown more simply by the CG analysis.  

\begin{figure}
\centering
\includegraphics[width=8.5cm]{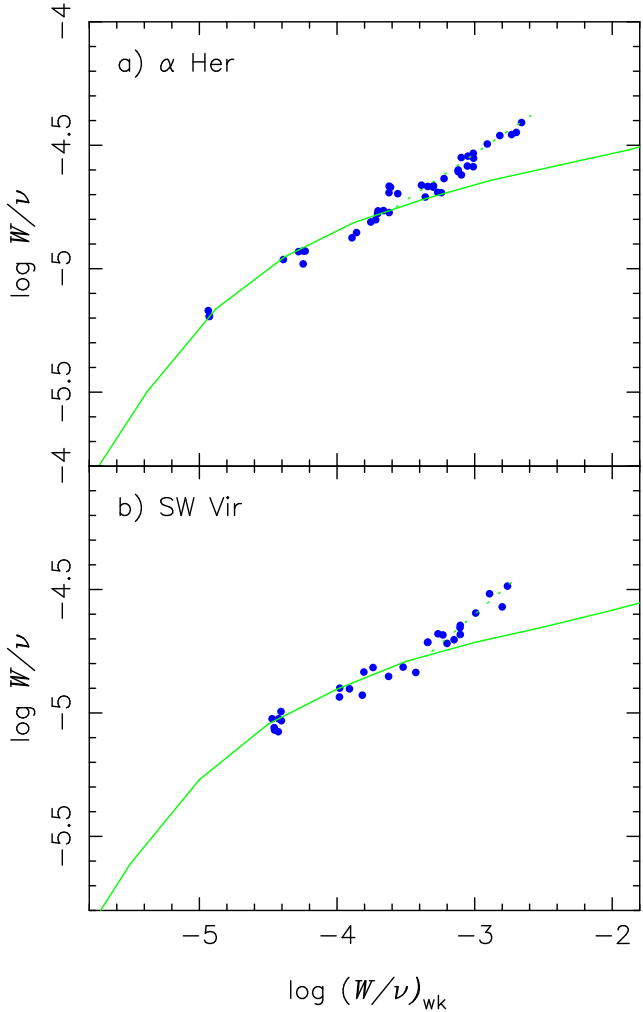}
\caption{
Empirical curves-of-growth of the OH fundamentals (filled circles) are 
compared with the theoretical ones (solid lines) based on the PP models 
and the microturbulent velocities given in Table 4. Note that the 
 intermediate-strength lines show systematically larger equivalent 
widths than those  predicted by the theoretical curves-of-growth.
{\bf a)}  $\alpha$ Her (M5Ib-II). {\bf b)} SW Vir (M7III:).
}
\label{Fig8.eps}
\end{figure}

 Based on the classical theory of line formation, the theory of 
curve-of-growth is well established (e.g. Uns\"old 1955).
However, such an upturn (or downfall) in the flat part of the 
curves-of-growth identified  in this subsection is quite unexpected, 
yet important to clarifying its origin.  There is probably little 
possibility of explaining these effects in the flat part of the 
curves-of-growth within the framework of the theory of line
formation in homogeneous photospheres. For example, 
for the low densities in the photospheres of giant stars, the upturn 
cannot be attributed to the development of the damping wings. 
Microturbulence would also not exhibit such an abrupt upturn as observed. 
For this reason, we propose that it is reasonable to interpret the 
unusual pattern of the curves-of-growth as evidence of the 
hybrid nature of the infrared spectra.

\subsection{Red supergiant stars}
 Because the CG method has been applied more often in the past, we 
can examine whether the anomalous behaviors of the intermediate-strength
lines  can be found in the curves-of-growth published by other
authors. First, we recall that Lambert et al.\,(1984) showed
in their Fig.\,2 a curve-of-growth of OH fundamental lines 
for M2 supergiant Betelgeuse and noted that the stronger lines cannot be 
fitted by the theoretical curve-of-growth. We 
note that the behavior of the stronger OH lines in Betelgeuse
is quite similar to that found in Arcturus as well as M giant
stars. Inspection of their Fig.\,2 indicates that the upturn
in the stronger OH lines also starts at about ${\rm log}\,W/\nu \approx
-4.75$ or at slightly weaker lines, and the deviation from the flat
part of the theoretical curve-of-growth is more significant in this
supergiant star than in the red giant stars that we have studied. 

Several possible origins 
of this unusual behavior of strong OH lines (or the
intermediate-strength  and the strong lines in our classification) 
in Betelgeuse were considered by Lambert et al. (1984), although
the problem  remains unexplored since then as far as we are aware.
Because of the similarity between the phenomenon in Betelgeuse and
 that in M giant stars, 
it is natural to interpret the magnificent upturn of
the curve-of-growth of Betelgeuse as being caused by an extensive MOLsphere 
in this red supergiant star.  

\subsection{Mira-type variables}
   The CG method has been extensively applied to the infrared
molecular lines in Mira-type variables by Hinkle and his colleagues.
First, Hinkle (1979) showed that the spectra of Mira variables are 
quite complicated in that a molecular line generally consists of 
multiple components with different temperatures and velocities. 
The CG method was extended to the H$_{2}$O ro-vibrational lines in R Leo
by Hinkle \& Barnes (1979), who found that two separate layers, warm
and cool, contribute to the H$_{2}$O spectrum. Their curve-of-growth
for the warm-component lines (their Fig.\,4) shows an upturn at
${\rm log}\,W/\nu \approx -4.75$ or at slightly weaker lines,
and the observed curve-of-growth for the cool-component lines (their
Fig.\,6) shows more clearly that the strong lines deviate significantly 
from the flat part of the theoretical curve-of-growth.

A time series of 32 high resolution FTS spectra of another Mira variable
$\chi$ Cyg was analyzed by Hinkle et al.(1982), who clarified the detail
 of the Mira variability  including the photospheric pulsation,
shock generation, and circumstellar structure with an
800K stationary layer discovered by them. We consider here only
their CO curves-of-growth of $\chi$ Cyg shown in their Fig.\,4:    
the curves-of-growth show significant changes in the shape of the 
strong line portion with phase and their gradients are often too
steep to be fitted to the flat part of the curve-of-growth expected
for the low gravity atmospheres typical of Mira variables. 
The gradients in some phases are indeed as steep as those expected for
the damping part of the curve-of-growth. We conclude that the peculiar
shapes of the curves-of-growth in the strong line portion are  rather  
general features in Mira variables. 

The anomalous shapes of the curves-of-growth in Mira variable stars 
are  not necessarily associated with  special problems of
variability, since  more or less similar phenomena are
 found in ordinary giants including Arcturus and many
M giants, and should  reflect some fundamental
problems of line formation in cool luminous stars in general.

\section{Line intensities, shifts, and shapes }  
   We so far discussed mainly equivalent widths and found that the 
strong lines show unusual behaviors difficult to explain by the 
classical theory of line formation. We further  examine whether 
 there is any peculiarity in the observed characteristics of the 
stronger lines compared to the weaker lines in Arcturus.

\subsection{Line intensities }
We plot the values of $ {\rm log}\,W/\nu$ against the lower excitation
potentials (LEPs) for the lines of the CO first overtone and fundamental
bands observed in Arcturus in Figs.\,9a \& 9b, respectively.  
The dotted lines show  the critical value of $ {\rm log}\,W/\nu \approx 
-4.75$. It is immediately clear that the lines stronger than the
critical value are limited to lines with LEP lower than
about 1.3 eV (overtone bands) or  1.5 eV (fundamental bands).
This means that the excess absorption in Arcturus should originate
in the region of low excitation but not necessarily be very low. 

\begin{figure}
\centering
\includegraphics[width=8.5cm]{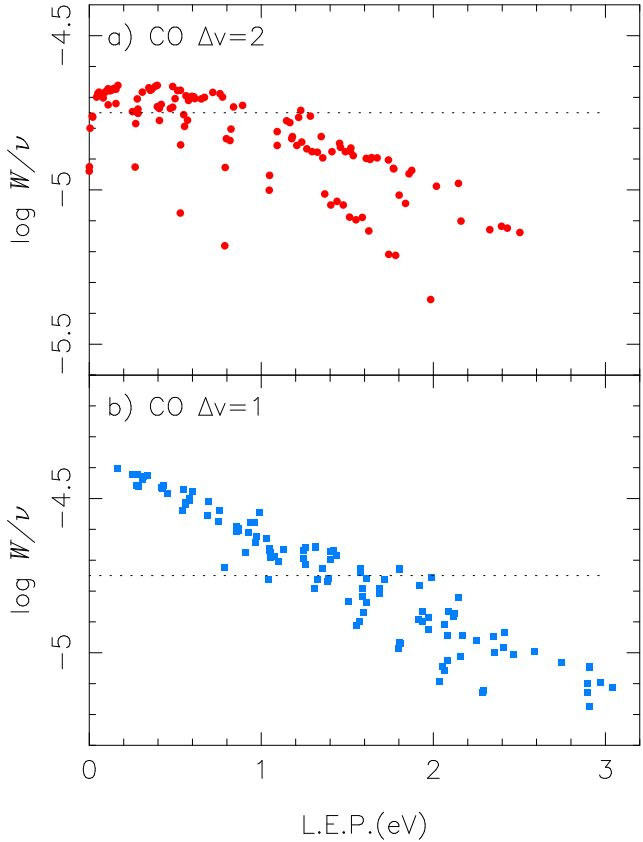}
\caption{
The values of $ {\rm log}\,W/\nu $ of CO lines observed
in Arcturus are plotted against the lower excitation potential (L.E.P). 
The critical value of $ {\rm log}\,W/\nu \approx -4.75$ is
indicated by the  dotted lines.
{\bf a)} The CO first  overtone bands.
{\bf b)} The CO fundamental bands.
}
\label{Fig9.eps}
\end{figure}

\begin{table} 
\centering
\caption{ Differential line-shifts between low (L) and high (H)
 excitation lines in radial velocities measured at the peak
absorption (P) and at the mid-point of FWHM (M). ME is the mean error
of the radial velocity. }
\vspace{-2mm}
\begin{tabular}{c c c c c}
\hline \hline
\noalign{\smallskip}
CO band  & $V_{\rm P}^{\rm H} - V_{\rm P }^{\rm L}$ (ME) & 
$V_{\rm M}^{\rm H} - V_{\rm M}^{\rm L}$ (ME) &  $N$  &  sp. ref.  \\
         &   (km\,s$^{-1}$) &   (km\,s$^{-1}$) &   & (Tab.\,1) \\
\hline
\noalign{\smallskip}
$\Delta v = 1$  &  0.61 (0.08) & 0.45 (0.07)  & 24  &  1 \\
                &  0.31 (0.07) & 0.52 (0.09) & 17  &  2 \\
                &  0.28 (0.09) & 0.37 (0.10) & 13  &  3 \\
                &  0.27 (0.08) & 0.24 (0.09) & 13  &  4 \\
                &  0.26 (0.07) & 0.35 (0.06) & 22  &  5 \\
                &  0.08 (0.03) & 0.18 (0.04) & 27  &  6 \\
\noalign{\smallskip}
$\Delta v = 2$  &  0.09 (0.01) & 0.07 (0.01)  & 117 &  7 \\
\noalign{\smallskip}
$\Delta v = 3$  &  0.01 (0.04)  & 0.00 (0.05) &  34 &  7 \\
\noalign{\smallskip}
\hline
\noalign{\smallskip}
\end{tabular}
\end{table}

\subsection{Line shifts}
We measure the shifts of individual lines and obtain the corresponding 
radial velocities for them. The different spectral regions
were observed at different epochs (see Table 1) and it is difficult  
to convert the radial velocities to a consistent absolute scale accurately 
because the reference laser and signal beams do not necessarily  pass through 
the FTS on parallel paths (Hinkle et al. 1995). However, we do not
need to know the radial velocities themselves for our purposes
and we measure differential line-shifts between high and low excitation
lines in each spectrum. We refer to the radial velocity  of a line 
as $V^{\rm H}$ or $V^{\rm L}$ according to whether its LEP is higher
or lower than 10,000 cm$^{-1}$ ($\approx 1.25$\,eV). We also measure
the shifts of a line at its peak absorption and mid-point of 
 FWHM, and distinguish them by suffixes P and M, respectively.

The resulting mean values of the differential line shifts in the 
peak position and the mid-point of FWHM are given in Cols.\,2 \& 3, 
respectively, of Table 5 with 
the mean errors (ME) of the radial velocities in  parenthesis. The
number of lines measured is given in Col.\,4 and the identification number
of the spectrum referred to in Table 1 is given in Col.\,5. Inspection
of Table 5 reveals that all the spectra exhibit  positive differential shifts
in $V^{\rm H} - V^{\rm L} $, indicating that the low excitation lines
(largely the intermediate-strength lines) are expanding slightly 
compared to the high excitation lines (mostly the weak lines). The
differential shifts, however, are smaller in the overtone bands than in
the fundamental bands. 

We also examine the line asymmetry defined by $V_{\rm P} - V_{\rm M }$
and the results are plotted against LEP in Figs.\,10a-d, separately
for different CO bands (a \& b: fundamentals by winter \& summer data,
respectively, c: first overtones, d: second overtones). The solid lines
represent linear fits to the data and it appears that the line asymmetry 
depends little on LEP.  The mean values of $V_{\rm P} - V_{\rm M }$ are
0.00, -0.08, -0.09, and -0.12 for a, b, c, and d, respectively. 
The result that the mean values of $V_{\rm P} - V_{\rm M }$
tend to be negative agrees well with our previous measurements for Arcturus
as well as for many M-giants (Tsuji 1991). Our analysis can also be
regarded as a highly simplified version of the bisector analysis, a
result of which on Arcturus for Fe I lines by Dravins (1987) indicates 
that $V_{\rm P} - V_{\rm M }$ should be negative from his Fig.\,1. 
We conclude that CO lines including those of fundamental bands show
only minor asymmetry if any.

\begin{figure}
\centering
\includegraphics[width=8.5cm]{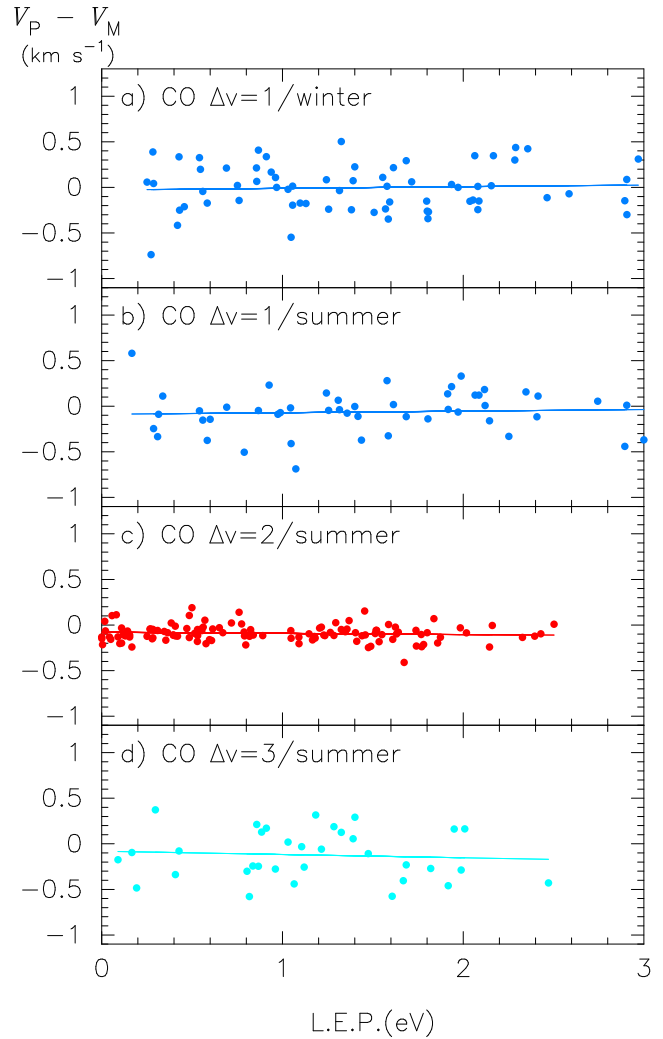}
\caption{
The line asymmetries defined by $V_{\rm P} - V_{\rm M}$, where
$V_{\rm P}$ and  $V_{\rm M}$ are the radial velocities at the peak
of absorption and at the mid-point of the half-maximum intensities,
respectively, are plotted against LEP (in eV). The solid lines
represent linear fits to the data. 
{\bf a)} The CO fundamental bands (winter data: sp. refs.1-4 in Table 1).
{\bf b)} The CO fundamental bands (summer data: sp. refs.5-6).
{\bf c)} The CO first  overtone bands (summer data: sp. ref.7).
{\bf d)} The CO second  overtone bands (summer data: sp. ref.7).
}
\label{Fig10.eps}
\end{figure}

\begin{figure}
\centering
\includegraphics[width=8.5cm]{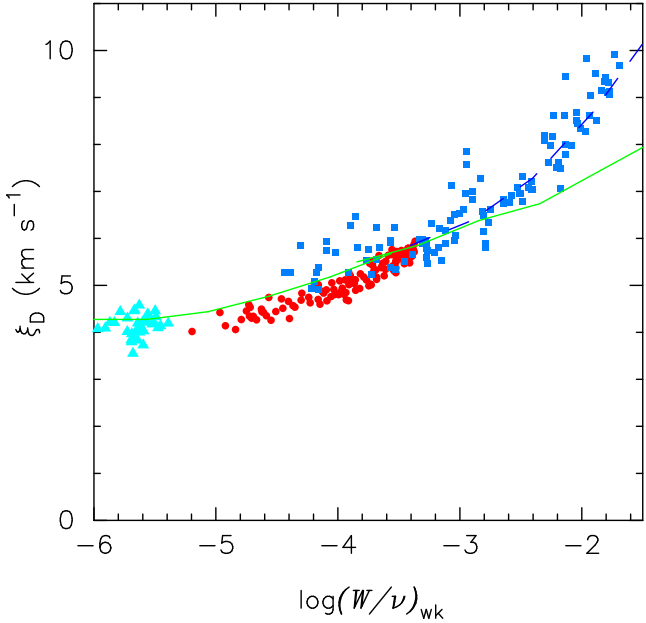}
\caption{
The Doppler velocity dispersions $\xi_{\rm D}$ (in km\,s$^{-1}$) of the 
CO lines observed in Arcturus are plotted against log\,$(W/\nu)_{\rm wk}$,
which can be a measure of the intrinsic line intensity. 
The CO fundamental, first,  and second  
overtone bands are represented by the filled squares, circles, and triangles,
respectively. The predicted $\xi_{\rm D}$ values based on the
photospheric model with $\xi_{\rm micro}$ =1.87\,km\,s$^{-1}$ and 
$\xi_{\rm macro}$ =3.47\,km\,s$^{-1}$ are shown by the solid lines.
The effect of molecular clouds with $\xi_{\rm micro}$ = 4.0\,km\,s$^{-1}$
in our ad-hoc model is shown by the dashed line.
}
\label{Fig11.eps}
\end{figure}

\subsection{Line broadening}
We  measure  FWHMs and line-depths, and the results are given in 
Table 2. The smoothing was applied to the atlas spectra and we
correct its effect on FWHMs with the Gaussian of FWHM = 1.25 x Res 
(Hinkle et al. 1995). The resulting intrinsic FWHM is converted to
the Doppler velocity dispersion $\xi_{\rm D}$ by FWHM = 
$2 \sqrt{\rm Ln2} (\nu/c) \xi_{\rm D}$. 
The resulting values of $\xi_{\rm D}$ are plotted in Fig.\,11
against the intrinsic line intensities represented by log\,$(W/\nu)_{\rm wk}$
defined by eqn.(1).  It is to be noted in Fig.\,11 that the lines of
CO fundamentals, first, and second overtones shown by the filled  squares, 
circles, and  triangles, respectively, form a continuous curve. At
first, we plotted observed $\xi_{\rm D}$ values against measured 
line-depths. However, plots of the overtone and fundamental lines
formed separate curves rather than a unified curve such as in Fig.\,11, 
and this may be because the line-depth cannot be a good measure of intrinsic 
line intensity especially if applied to different spectral regions.

We assume that the lines of the CO second overtones and relatively
weak lines of the first overtones should  originate in the
photosphere from our analysis outlined in Sect.\,3.1. The line-widths 
of these lines then can be interpreted as being photospheric in origin. We 
extrapolate the measured Doppler velocities to the weak-line limit 
in Fig.\,11 and find the limiting $\xi_{\rm D}$ value of 4.07
km\,s$^{-1}$. This result cannot be explained by the
microturbulent velocity of 1.87 km\,s$^{-1}$ and thermal velocity
dispersion of $\xi_{\rm th} \approx$ 1 km\,s$^{-1}$. We conclude that 
an additional broadening with velocity dispersion of 
$({ \xi_{\rm D}^2  - \xi_{\rm th}^2 - \xi_{\rm micro}^2  })^{1/2} \approx$ 
3.47 km\,s$^{-1}$ is required and we attribute its origin to large-scale
photospheric motions, which are referred to as macroturbulence.

It is by no means clear, however, if all the results shown in Fig.\,11 
can be understood with the micro and macroturbulent velocities noted 
above. We examine this problem by evaluating the Doppler velocities for
the fictitious lines used to generate the theoretical curves-of-growth
in Sect.\,4.1. To each computed line profile based on the photospheric
model (model a in Fig.\,2) with $\xi_{\rm micro}$ =1.87\,km\,s$^{-1}$
throughout the photosphere, we apply the broadening caused by macroturbulence 
with velocity dispersion of $\xi_{\rm macro}$ =3.47\,km\,s$^{-1}$,
and measure the Doppler velocity as we did for the observed spectra. 
The results are shown in Fig.\,11 by the solid lines, of which
the weak and strong line parts are  based on the fictitious lines of 
the CO first overtones and fundamentals, respectively. The weak and
strong line parts join at log\,$(W/\nu)_{\rm wk} \approx -3.5$.
The predicted and observed Doppler velocities roughly agree  for
most lines with log\,$(W/\nu)_{\rm wk} \la -3.0$, although
further fine tuning in the thermal and/or velocity structures
should be required to have a perfect fit.

We know that the lines of the CO fundamentals do not originate in
the photosphere alone.  The observed Doppler velocities in fact appear to be
much higher than the predicted ones for the stronger CO lines with
log\,$(W/\nu)_{\rm wk} > -3.0$ in Fig.\,11. Clearly, the
stronger CO lines require additional broadening other than the 
micro and macroturbulence so far considered.
The stronger lines of the CO fundamentals, which 
can be explained  neither by the RE photospheric model nor by
the hot chromospheric model (Fig.\,5b), are showing evidence for a  
cool constituent responsible to the
strong absorption.  The additional broadening of the
stronger fundamental lines shown in Fig.\,11 is consistent with 
this conclusion from our CG analysis, and the cool constituent 
responsible for  additional absorption should  be related to 
significantly large turbulence (further discussed in Sect.\,6.2).
 
\section{Discussion}
Some  observational data for Arcturus and  other cool luminous stars 
cannot be explained by our  present understanding of stellar
atmospheres. Although we have no solution on this difficulty at present, 
we examine some possibilities.   

\subsection{ The case of Arcturus }
   
    A difficulty in the spectrum of Arcturus was highlighted  by the 
detailed  line-by-line analysis, which exhibited excess absorption in 
the lines of $ {\rm log}\,W/\nu > -4.75$ (Figs.\,3 \& 4).
The specific lines that show excess absorption originate in levels 
of relatively low excitation potentials but as high as 1 eV or 
higher (Fig.\,9). This fact implies that the excess absorption 
should be formed in an environment at relatively cool but not extremely 
cool. The problem is also  evident in the curves-of-growth, which show
an unexpected upturn in their flat part at about
$ {\rm log}\,W/\nu \approx -4.75$ (Fig.\,5). This sudden upturn is
difficult to understand within the framework of the line formation
theory in a homogeneous photosphere, and a possible
explanation may require an additional constituent other than
the hot chromosphere, which tends to weaken  rather than
to strengthen the CO lines.

In the outer atmosphere of Arcturus, 
the possible presence of a cool component referred 
to as a CO-mosphere was suggested to explain the strong CO lines by 
Wiedemann et al. (1994), who considered a thermal bifurcation model
consisting of hot and cool components. 
The cool component was proposed to be generated by the molecular cooling 
caused by CO and other molecules. This molecular cooling was, however,
already included in constructing 
the RE model photosphere and, for this reason, the surface
temperature of our model employed in this study is  as low as 
$\approx 2300$\,K at log\,$\tau_{0} = 10^{-6}$ 
(Fig.\,2).  However, as we know already, this RE model photosphere
of very low surface temperature could not explain 
the excess absorption of the strong CO fundamental lines at all
(Figs.\,4 \& 5b).   

For the reason outlined above, we propose that
an additional constituent other than the cool and hot components  
shown in Fig.\,2 should be introduced. The presence of such a
third constituent would be consistent with the large Doppler velocities of
the strong CO fundamental lines as noted in Fig.\,11. However,
the strong CO fundamental lines appear to exhibit only minor line-asymmetry
 and only small outflow motion relative to the CO weaker 
lines that possibly originate in the photosphere, as shown in Sect.\,5.2.

From the empirical data summarized above, we suggest that
 molecular condensations may be formed within the atmosphere of
Arcturus. The molecular condensations may not necessarily
form a separate structure but may be floating in the atmosphere
as molecular clouds. These molecular clouds then show only small  
motions relative to the surrounding atmosphere from which the
clouds are formed, and the CO lines should not necessarily exhibit
asymmetry nor high outflow velocity. The turbulent velocities in the
clouds, however, can be high due to possible dynamical effects
associated with the cloud formation, and the large Doppler velocities 
of the strong CO fundamental lines (Fig.\,11) can be consistent with this
scenario.

  A problem is how the additional material needed for cloud formation 
can be made available and how the molecular clouds can be formed.
For this purpose, radiation and turbulent pressures are insufficient, 
as is clear from the fact that extended hydrostatic models included
such effects resulted in  only very low density extended photosphere 
(Sect.\,2.4 and Tsuji 2006). One possibility is that the atmospheres 
of red giant stars are not so clear as assumed in the classical RE 
model photospheres but may contain some additional material supplied 
by surface activities of various kinds (Sect.\,6.4). Once density 
inhomogeneity appears in the huge atmosphere of red giant stars, 
it may grow further by the thermal instability triggered by the 
cooling due to the strong infrared radiation by molecular bands 
as suggested, for example, by Muchmore et al. (1987).
The molecular cooling has already been included in the hydrostatic RE
model photospheres, but this mechanism can play a more important role
in cloud formation if applied to the gaseous material in free space.
Cuntz \& Muchmore (1994) demonstrated that domains of radiative
instabilities caused by CO and SiO cooling exist within the outer 
atmospheres of red (super)giant stars, and molecular clouds can be 
formed in these instability islands. 

\subsection{ An ad-hoc model}
We propose the presence of molecular clouds in the
atmosphere of Arcturus to explain the available observed data
presented in Sects.\,3 - 5. At present, we have no precise idea 
about the location, temperature, or density of the clouds. However, 
we now try some numerical exercises to have some assessments on the 
nature of the clouds by considering an ad-hoc model. 
We first assume that the clouds may be formed in the upper
atmosphere where strong lines are formed and temperatures can 
in general be low. Also, if many clouds are formed to have definite
observational effects, the net effects of them can be approximated
by a  shell consisting  of many overlapping clouds, and we apply 
the MOLsphere model that we assumed in the case of red supergiant 
stars (Tsuji 2006). This  does not necessarily imply that a separate 
``sphere'' exists in the outer part of the atmosphere but we use 
this model only for convenience in numerical analysis.

\begin{figure}
\centering
\includegraphics[width=9.0cm]{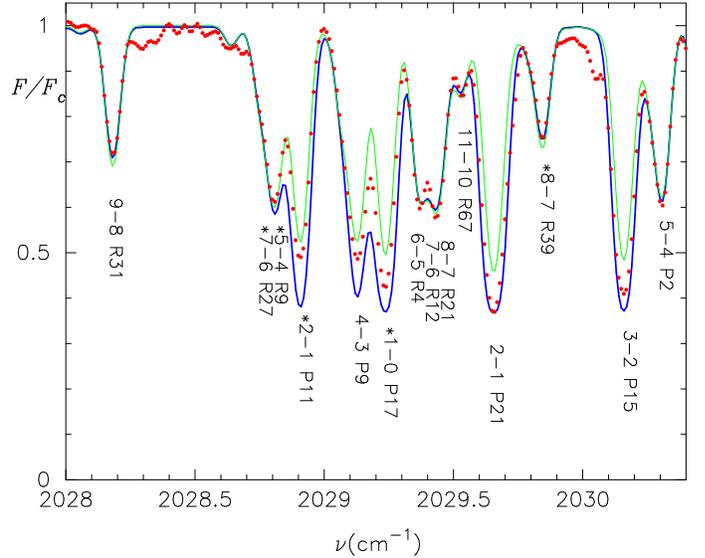}
\caption{
Predicted spectra based on the RE model
photosphere (thin line) and the ad-hoc model 
discussed in the text (thick line) are compared with  the observed 
spectrum of Arcturus (filled circles). 
The line identifications given by Hinkle et al. (1995) are
reproduced: $^{13}$C$^{16}$O lines are marked with asterisks
and  $^{12}$C$^{16}$O lines without.
}
\label{Fig12.eps}
\end{figure}

\begin{figure}
\centering
\includegraphics[width=8.5cm]{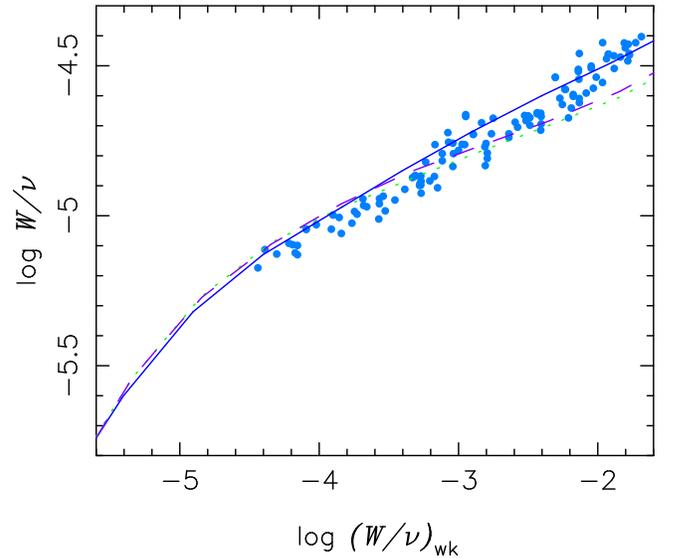}
\caption{
Predicted theoretical curve-of-growth for our ad-hoc model 
(solid line) is compared
with the observed data on CO fundamental lines (filled
circles) and the theoretical curve-of-growth for the RE
photospheric model (dotted line) reproduced from Fig.\,5b.
Predicted theoretical curve-of-growth for another ad-hoc model 
in which temperatures in the surface layers are reduced by as much as
250\,K is shown by the dashed line (as for detail, see the text).
}
\label{Fig13.eps}
\end{figure}

We start from an extended photosphere in radiative and 
hydrodynamical equilibria (SS model) starting integration from
log\,$\tau_{0} = -9.0$ (model b in Fig.\,2).
The geometrical extension $R_{\rm out}$ depends on the turbulent 
pressure and we assume a turbulent velocity $\xi_{\rm Mol}$ as a 
free parameter. Our extended RE model with  $\xi_{\rm Mol} = 
6$\,km\,s$^{-1}$ (log\,$\tau_{0} = -9.0$, model b in Fig.\,2) 
produced $R_{\rm out} = 28.33\,R_{\odot} = 1.11\,R_{*} $.
We now make ad-hoc assumptions of a uniform temperature of
$T_{\rm Mol}$ and a column density of $N_{\rm Mol}$ in the layers
above a critical depth of log\,$\tau_{0}^{\rm cr}$ at a geometrical
radius $R_{\rm in}$. The resulting
model is in neither radiative nor hydrostatic equilibrium.
Then, we take log\,$\tau_{0}^{\rm cr} = -6.0$ where 
$R_{\rm in} = 26.57\,R_{\odot} = 1.04\,R_{*} $.  
We note that this model is essentially identical 
to introducing a shell of given inner and outer radii, turbulent velocity, 
temperature, and density. We use this ad-hoc model simply because 
our spectral synthesis code can be applied directly to such a model and
the photosphere can be included automatically as a boundary condition.   

We try several combinations of $T_{\rm Mol}$ and $N_{\rm Mol}$
for the shell with inner and outer boundaries at $1.04\,R_{*} $ 
and $1.11\,R_{*} $, respectively.
We have assumed the turbulent velocity of  6\,km\,s$^{-1}$ in
modeling the extended photosphere, but this was simply as a means 
by which to generate our ad-hoc model based on the extended RE model. 
We must determine microturbulent velocity $\xi_{\rm micro}^{\rm Mol}$ in 
the MOLsphere anew to be consistent with  observations. For this
purpose, we try several values for $\xi_{\rm micro}^{\rm Mol}$ and
examine if the observed Doppler velocities of the strong CO lines in 
Fig.\,11 can be reproduced with our fictitious lines of the CO
fundamentals used in Sect.\,5.3. This analysis is done iteratively with 
that of the observed spectrum of CO fundamentals represented by Fig.\,12 
(filled circles). For this purpose, we now compute a synthetic spectrum 
of CO fundamentals corresponding to  Fig.\,12 by the use of our ad-hoc 
model with $\xi_{\rm micro}^{\rm Mol}$ in the MOLsphere and $\xi_{\rm micro} = 
1.87$\,km\,s$^{-1}$ in the RE photosphere included as the boundary 
condition. Some lines of $^{13}$CO are included  and  we assume a 
$^{12}$C/$^{13}$C ratio of 7 (Hinkle 1976). The resultant synthetic
spectrum is convolved with the smoothing function of the atlas (Gaussian 
with FWHM = 1.25 x Res) and the macroturbulent broadening of the
velocity dispersion $\xi_{\rm macro} = 3.47$\,km\,s$^{-1}$ (Sect.\,5.3).

The results of our trial and error are shown in Fig.\,11 for the
Doppler velocity and in Fig.\,12 for the synthetic spectrum. 
The parameters that provide  reasonable fits are found to be:
$\xi_{\rm micro}^{\rm Mol}$ = 4.0\,km\,s$^{-1}$, $N_{\rm Mol}({\rm CO}) 
=$ 5 x $10^{+19}$\,cm$^{-2}$
\footnote{ With the C/H ratio of 10$^{-4}$, this CO column density
implies H column density of 5 x $10^{+23}$\,cm$^{-2}$ or mass column
density of about 1\,g\,cm$^{-2}$, which is about the same as that
above the temperature minimum in the empirical chromospheric model
(Ayres \& Linsky 1975).}, 
and $T_{\rm Mol} = 2000$\,K.
With these parameters, the large Doppler velocities of the stronger CO 
lines are reproduced as shown by the dashed line in Fig.\,11.   
As shown in Fig.\,12, the observed strong CO lines (filled circles)
can be reproduced approximately by the prediction based on our 
ad-hoc model (thick line), but are too deep to be accounted for by 
the  predicted spectrum based on the RE model photosphere alone 
(thin line).  On the other hand, the observed weaker CO lines 
(e.g. 9-8 $R$31, $^{*}$8-7 $R$39) are slightly shallower than the 
predictions based on the RE photospheric model (thin line), in a way 
consistent with the CG analysis (Fig.\,5b), but can be fitted by the 
predictions based on our ad-hoc model (thick line). 

We also compute theoretical curve-of growth for our ad-hoc model
with the same parameters used in the computation of the synthetic
spectrum and the Doppler velocity. The result is shown by the solid 
line in Fig.\,13. For comparison, we reproduce the predicted CG for the
RE photospheric model (dotted lines)  together with the observed data 
(filled circles) from Fig.\,5b.
Now, the basic features of the observed data are that the relatively weak
lines are  below the theoretical CG for RE photospheric model and the 
stronger lines are above it, which can roughly be accounted for by the 
theoretical CG including the effect of molecular clouds (solid line).
The fits, however, are not so good for lines around log\,$W/\nu \approx
-5.0$ and further refinement of our model is needed.

For the result that the weaker CO fundamental lines tend to be below 
the theoretical curve-of-growth based on the RE photospheric model 
(Fig.\,5b),  we recalled in Sect.\,4.1 the  possibility that the EWs 
can be reduced slightly  by the emission caused by the chromospheric 
temperature inversion (Heaseley et al. 1978). An alternative possibility 
to resolve this issue, however,  is that the reduction in EWs may be 
caused by the thermal
emission of the molecular clouds as shown in Fig.\,12. We  prefer the
possibility of molecular clouds rather than chrompsphere, because the 
molecular cloud model provides a consistent interpretation of the weaker and 
stronger CO fundamental lines, while the chromospheric effect is supposed to 
be rather minor in Arcturus if present at all (Wiedemann et al. 1994). 

\begin{figure}
\centering
\includegraphics[width=8.5cm]{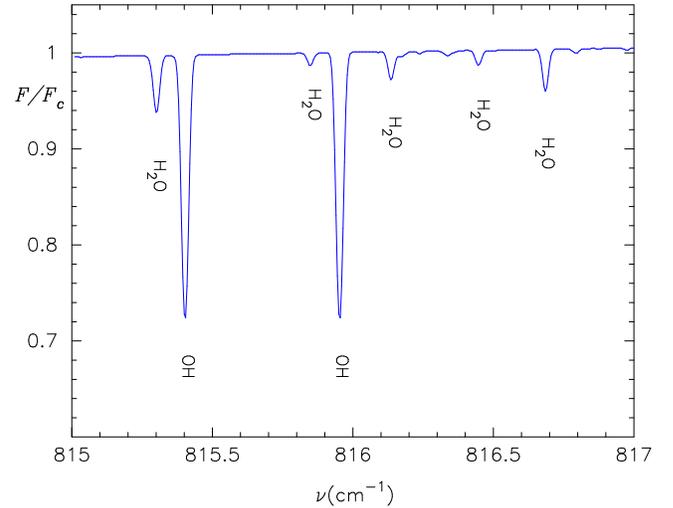}
\caption{
Predicted spectrum of the pure rotation lines of OH and
H$_{2}$O  based on an ad-hoc model 
discussed in the text. OH lines are mostly formed in the photosphere
while H$_2$O lines in the molecular clouds or the MOLsphere. As to the 
observed spectrum of Arcturus, see Fig.1 of Ryde et al.(2002).
}
\label{Fig14.eps}
\end{figure}
 
The case discussed above is probably neither the optimal nor unique 
solution,  but we show only an example of how the observed 
deep cores of the CO fundamental lines can be accounted for
with the molecular clouds, whose total CO column density is about
5 x $10^{+19}$\,cm$^{-2}$. We do not think that it is useful 
to explore such an
ad-hoc model in more detail. We note, however, that the extension of 
the shell cannot be very large, since strong CO lines will appear as 
emission lines for $R_{\rm in}$ and/or $R_{\rm out} $ larger than 
those assumed above (the weaker lines already contribute to emission as
noted in the preceding paragraph)
\footnote{This example that the
molecular clouds produce either emission or absorption depending
on the line-strengths (and possibly on other parameters that define the shell)
lends a supporting argument for our interpretation of  either strengthening
(e.g. $\beta$ Peg and $\alpha$ Her in Fig.\,6) or weakening (e.g. $\rho$
Per and SW Vir in Fig.\,7) of the intermediate-strength lines by the
contribution of the MOLsphere (for further details, see Tsuji 2008).}, 
so long as we  assume a spherically symmetric and homogeneous shell. 

We also examine whether the unexpected detection of pure rotation  lines 
of H$_{2}$O by Ryde et al.(2002) in Arcturus can be interpreted
in the same way. For this purpose, we retain all the features in our
ad-hoc model discussed above except for the column density.  We
assume again the microturbulent velocities of $\xi_{\rm micro}^{\rm Mol}$ 
= 4.0\,km\,s$^{-1}$ in the MOLsphere and $\xi_{\rm micro}$ 
= 1.87\,km\,s$^{-1}$ in the photosphere, and apply linelists of H$_2$O 
(Partridge \& Schwenke 1997) and OH (Jacquinet-Husson et al. 1999).
An example of the predicted spectrum with  $N_{\rm Mol}({\rm H_{2}O}) =$ 
2 x $10^{+17}$\,cm$^{-2}$ is shown by the solid line in Fig.\,14.
The observed H$_{2}$O lines (Fig.\,1 of Ryde et al. 2002) can  
be accounted  for roughly by our ad-hoc model. 

This is certainly not a unique solution: For example, Ryde et al.\,(2002) 
showed that their observation of H$_{2}$O lines in Arcturus can be 
explained if the surface temperatures of the RE model can be lowered 
by about 300\,K. A possibility of such a surface cooling was predicted 
by non-LTE models of Arcturus (Short \& Hauschildt 2003). The largest 
effect was shown for a plane-parallel case including Fe, Ti, and light 
elements in NLTE, and the surface temperature was lowered 
by about 250\,K compared to the corresponding LTE model. 
We introduce the lowerings beginning with 250\,K at the surface, as
shown in  their Fig.\,1, into our LTE model, and
solve chemical and hydrostatic equilibria for the modified
temperature structure. We use the resulting model to computing CO lines
and construct  curve-of-growth as we did for our ad-hoc model.
We assume microturbulent velocity of 1.87\,km\,s$^{-1}$ 
throughout the photosphere as in our original photospheric model.
The result is shown by the dashed line in Fig.\,13.

Inspection of Fig.\,13 reveals that the CO lines can also be strengthened  
somewhat by the NLTE cooling in the surface layers. 
The strengthening is rather modest but it can be larger if
larger microturbulent velocity is assumed in the surface layers
as may be suggested from Fig.\,11. However, it is needed to find
a  velocity structure that explains consistently all the observed
features such as shown in Figs.\,11 - 13, and  such a possibility
should hopefully be explored. At present, however, our ad-hoc 
model is consistent with all these observed features as shown in 
Figs.\,11 - 13.  Also, relatively weak lines remain unchanged so far as 
photospheric models are applied  and hence the weakening of the relatively 
weak lines cannot be explained by such models. By our ad-hoc model,
however, the weakening can be explained as due to the thermal emission of the 
molecular clouds.

The pure rotation lines of H$_2$O at 12\,$\mu$m were also observed in 
red supergiants such as $\alpha$ Ori (Ryde et al. 2006a) and  $\mu$ 
Cep (Ryde et al. 2006b) in absorption,  and it was difficult to 
understand why they were not detected in emission if water molecules 
reside in the extended outer envelope. Possibly for this reason, Ryde 
et al. (2006a,b) suggested that the water lines should originate 
in the photosphere. However, we know that the H$_2$O $\nu_2$ 
fundamentals at 6\,$\mu$m as well as the pure rotation lines at 
40 $\mu$m  appear in emission in $\mu$ Cep (Tsuji 2000b). A 
possible way to resolve this dilemma can be our proposal that dust 
grains expected to condense in the envelope provide a continuum background
(at around 12\,$\mu$m) against which the H$_2$O lines can be seen in 
absorption. However, we remained concerned that  the dust grains may 
not able to be accommodated in the envelope, where radiation pressure 
can expel the dust grains (see note added in proof in Tsuji 2006). 
However, a  detailed analysis of the
infrared spectrum of $\alpha$ Ori possibly detected
amorphous alumina in the extended envelope (Verhoelst et al. 2006).
Also, clean dust grains such as alumina and silicate were shown not to 
suffer the effect of radiation pressure in full (Woitke 2006; 
H\"ofner \& Andersen 2007). Then, the 12\,$\mu$m water 
lines in absorption  can also be produced by  
molecular clouds in the outer envelope of red supergiants. 
In other words, the 12\,$\mu$m H$_2$O absorption lines can 
consistently be understood as due to originating in molecular clouds 
in Arcturus as well as in red supergiants.

\begin{figure}
\centering
\includegraphics[width=9.0cm]{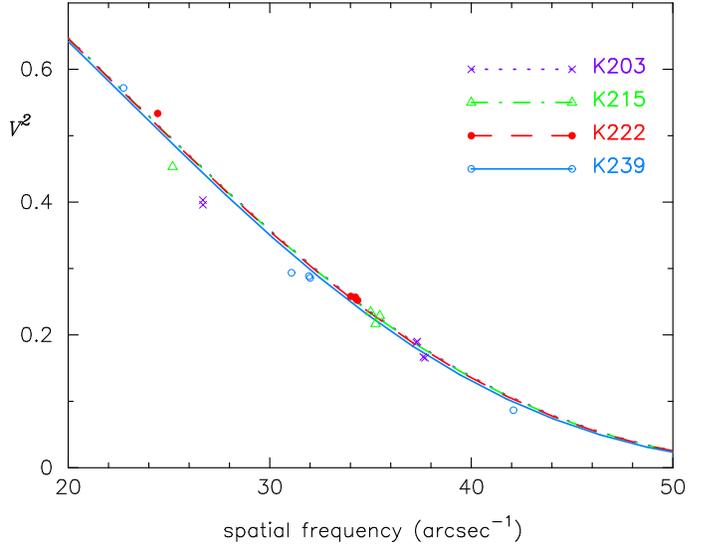}
\caption{
Predicted visibilities squared  based on the ad-hoc model 
discussed in the text are compared with the observed
visibilities squared by Verhoelst (2005). The results for different
filter bands are shown by different symbols and lines as 
noted in the figure.
Predicted visibilities squared  based on the RE model photosphere
show only minor differences from those based on the ad-hoc model.
}
\label{Fig15.eps}
\end{figure}

Finally, we examine whether our ad-hoc model can be constrained further  
by visibility measurements in four narrow-band filters, which
show the wavelength dependence of the diameter of Arcturus (Verhoelst 
et al. 2005). Interferometric observations were completed in 
four  narrow-band filters as for red supergiant star $\mu$ Cep
(Perrin et al. 2005), and predictions of the visibilities based on
models are carried out in exactly the same way as for
$\mu$ Cep (Tsuji 2006). We use the stellar parameters in Table 3
and assume  the angular diameter of the stellar photosphere to be 21.05 
arcsec (Lacour et al. 2008) throughout.

The predicted visibilities squared based on our ad-hoc model 
are compared with the observed data in Fig.\,15. First, we note
that the results for  the four bands show little difference, but
the predicted visibility squared for $K$239 band, which includes CO
first overtone bands, indicates that the diameter for this band may
be extended slightly. Second, some visibility data cannot be fitted
well with the predicted visibilities, as noted already by Verhoelst 
et al. (2005). The poor fittings are not necessarily associated with
the presence of molecular absorption in the filter bands and we
conclude that this difficulty cannot be resolved by a simple
model of the outer atmosphere. 

The results for the RE photospheric model (SS model) are very similar 
to the results for the ad-hoc model.  Since the extension of our
ad-hoc model  and hence its effect on visibilities are rather minor,
the results in Fig.\,15 may be evidence neither for nor against
our ad-hoc model. The results of  $\chi$-square test for the fits
are: $\chi^{2}/(N-1)$ = 11.60, 12.40, 1.71, and 5.23  for
$K$203, $K$215, $K$222, and $K$238 bands, respectively ($N$ is 
the number of observed data points) for the ad-hoc model, and 
$\chi^{2}/(N-1)$ = 10.93, 11.82, 2.13, and 5.13 for $K$203, 
$K$215, $K$222, and $K$238 bands, respectively, for the RE 
photospheric model (SS).
   
\subsection{From  Arcturus to  cool luminous stars }
We noticed excess absorption in CO lines of  M giant stars, first in the
low excitation strong lines in late M giants. The excess absorption was
difficult to interpret as being  photospheric in origin, and we  
proposed instead that it originates from cool molecular layers (which 
are however warmer than  the expanding circumstellar envelope) referred
to as a quasi-static molecular dissociation zone (Tsuji 1988).  
We then  found more or less similar unusual phenomenon not
only in the low excitation strong lines but also in the
intermediate-strength lines of CO and OH  with LEP as high as 2 eV
in dozens of  M-giant stars including K5 
giant $\alpha$ Tau (Tsuji 2008). We have shown in this paper that this
phenomenon can be recognized more simply by the curves-of-growth
(Figs.\,6-8). We have found in this paper that the same phenomenon 
appears in the early K giant $\alpha$ Boo and that the phenomenon can be
seen more clearly in the lines of the CO fundamentals. 
Thus we have had to recognize that the unusual behavior of 
molecular lines is probably  a more general phenomenon, which may be related
to some basic property of cool stellar atmospheres.

In M giant stars, the intermediate-strength lines of unusual 
behavior  showed little relative motion ($\la 1$\,km\,s$^{-1}$) 
compared to the weak lines (Tsuji 1991), and they may also originate 
in molecular clouds formed within the outer atmospheres. We recall, 
however, that some strong low excitation lines (log\,$W/\nu > -4.4$ and 
LEP $\la$ 0.5 eV) in the cooler M giants showed relative motions of 
a few km\,s$^{-1}$ (Tsuji 1988), which can be explained by some 
molecular clouds in the outer part (and hence cooler) beginning to decouple  
dynamically from the main bodies of the molecular clouds.  
We may  refer to an aggregation of the molecular clouds, including those 
with low relative motions, as a MOLsphere for simplicity in K and M 
giants as well as in red supergiants throughout.

The spatial extension of the MOLsphere has already been demonstrated
directly by observations of red supergiant stars with spatial interferometry 
(e.g. Perrin et al. 2005, 2007). The case of red supergiants shows
a marked contrast to the case of Arcturus for which interferometry
shows little evidence for an extended MOLsphere (Sect.\,6.2). 
Clearly the details of the MOLsphere may differ significantly
in different types of stars. The case of red supergiants may
represent a more advanced phase in the evolution of molecular clouds
discussed in the preceding paragraph for cooler M giant stars.
 The molecular clouds probably expand easily under the
lower gravities of the outer atmosphere of supergiant stars.

In the case of Mira variables, gaseous matter can be easily levitated 
by pulsations and/or by shocks, and the nature of molecular clouds
or MOLsphere may not necessarily be the same as in non-Mira variable stars.
Anyhow the presence of the molecular shells in the outer atmospheres of
Mira variables has clearly been shown by  interferometric observations  
(e.g. Perrin et  al. 2004b), by spectro-imaging 
(Le Bouquim et al. 2009), and by a  differential spectral imaging
with an adaptive optical system (Takami et al. 2009). 
The spectra of Mira variables have detected
multiple velocity components, and the different layers related
to different velocities have anomalous curves-of-growth
(Sect.\,4.4). This fact suggests that the molecular clouds may
form in each layer of different velocities (and hence at different
locations), and detailed analyses of spectral and interferometric 
observations of Mira variables will shed further light on the 
molecular clouds in the outer atmosphere of cool luminous stars. 

\subsection{Formation of molecular clouds and origin of mass-loss}

We have considered a possibility that the molecular clouds or MOLsphere   
may exist through early K giants to coolest M giant stars. 
The major problem is  how a large amount 
of gaseous material can be supplied to the outer atmosphere extended 
to a few stellar radii in the case of cooler (super)giant stars. 
In the case of Arcturus, the clouds may be formed not so far from 
the photosphere but the problem of how gaseous material can be transported 
there is the same. We have assumed a various kinds of surface activities 
(Sect.\,6.1), but their detail is  not specified. We now return to this 
subject for the case of non-Mira variable stars.

For this purpose, we recall that the surface activities are driven
mainly by magnet-hydrodynamical (MHD) processes in stellar envelopes.
For example, reappearance of dynamo activity may be due to  convection 
in the sheared rotation layers 
generated between the spinning-up contracting core and spinning-down 
expanding envelope as the star evolves from the subgiant to the giant 
phase  (Uchida \& Bappu 1982). This phenomenon, which occurs in 
the red giant phase, can explain the revival of the surface 
chromospheric activity due to the regenerated magnetic field, 
and lead to some material being injected into the 
outer atmosphere by associated surface activities.
 
A detailed MHD simulation of stellar winds 
in red giant stars starting from the photosphere in open magnetic 
field regions was performed by Suzuki (2007), who showed that 
the perturbations from the surface convection excite waves that 
propagate upward. 
An important result of Suzuki's simulation is that a nearly static
region is formed at several stellar radii above the photosphere. 
The gaseous matter levitated by each MHD process will accumulate  
around the nearly static region  and  grow to become a reservoir of 
abundant gaseous material. This reservoir can be a seat of molecular 
cloud formation, in which the molecular cooling mechanism noted in 
Sect.\,6.1 will also play a role. Such a reservoir has  
been detected observationally:
We may identify the MOLsphere or the aggregation of molecular clouds
formed in the outer atmospheres of early K to late M giants 
(Sect.\,6.3) with the nearly static reservoir predicted by the
MHD simulation. 

Another important point is that the stellar winds are effectively 
accelerated from the nearly static region, or the reservoir, located at
several stellar radii above the photosphere. Suzuki (2007) noted 
that this result explains why the observed flow velocity of 
the winds is rather low at about the escape velocity for several 
stellar radii above the photosphere. Thus, we conclude that the 
origin of stellar mass-loss in red giant stars has basically been 
resolved  from theory and observations consistently. This result 
is particularly welcome in view of a lack of successful explanation 
for mass-loss mechanism in red giant stars, since  previously 
proposed dust-driven mechanism was found not to be efficient 
enough to drive the mass-loss from oxygen-rich giant stars 
(Woitke 2006; H\"ofner \& Andersen 2007).
  
Generally, however, hydrodynamical processes including wave and shock 
formations result in rather violent variations in thermal and velocity 
structures of the outer atmosphere, which may have some observational 
consequences.  On the other hand, violent phenomena  are by no means 
evident in the observed data that we have examined,  and further detailed 
consideration is needed to reconcile theories more successfully with 
observations. For example, each hydrodynamical event may have little 
observable effect if matter involved in a single event is rather 
small and only the accumulated effect may have observable effects. 
 
\section{Concluding remarks}
We acknowledge the importance of observational data to our present 
work, in particular the excellent atlas of Arcturus (Hinkle et al. 
1995). We were able to measure more than 100 lines
of the CO first overtone bands based on the atlas
spectrum in which telluric features had been effectively removed
and, as a result, we found unexpected patterns in both 
our LL (Fig.\,3) and CG (Fig.\,5a) analyses of CO lines.
On the other hand, we could measure only about 20 lines
in a FTS spectrum from the KPNO archives, selecting only the 
lines undisturbed by the atmospheric lines and, as a result, 
we overlooked the anomalous behavior of the stronger CO lines 
(e.g. Fig.\,1a in Tsuji 1986).

More importantly, many lines of the CO fundamentals have been made
available by the atlas and we confirmed that the CO fundamental
lines provide very excellent insight into the cool molecular constituent 
of Arcturus (see e.g. Figs.\,4, 5b, 11, \& 12). The CO fundamental lines 
should certainly provide an excellent means of probing  molecular clouds in 
cooler (super)giant stars, and analyses of the CO fundamental lines should 
hopefully be extended to many cool luminous stars, despite the
well known difficulty caused by the strong atmospheric absorption in the
$M$ band region where the CO fundamental lines are observed.
High resolution spectroscopy of the $M$ band region from
space would be ideal, although we are unaware of any such plans.  

In our interpretation of the spectra, a major restriction is 
to have assumed LTE throughout. We have referred mainly to the
detailed non-LTE analysis of CO line formation by Ayres \&
Wiedemann (1989) and Wiedemann \& Ayres (1991). Departure from LTE 
should be certainly more important in the outer atmosphere
and it is  desirable that more or less similar NLTE analysis 
can be extended to the CO line formation in molecular clouds. 
Departure from LTE should also appear in other processes including
atoms and other molecules. An example for some atoms
(Short \& Hauschildt  2003) was already considered (Sect.\,6.2), 
although the problem continues to exist because of 
the difficulty in UV fluxes having significant effects on excitation 
and ionization (Short \& Hauschildt  2009). It is a formidable task
to take all the important processes consistently in a NLTE analysis
and we hope further progress in this field.  
  
Also,  chemical processes in the molecular clouds    
should most probably not follow LTE. We estimated CO and  H$_2$O
column densities in the molecular clouds and H$_2$O/CO ratio appeared
to be 4.0 x 10$^{-3}$ (Sect.\,6.2). At $T = 2000$\,K, this ratio 
corresponds to an equilibrium value at log\,$P_{\rm g} = -1.63$, which 
may not be very unreasonable for $P_{\rm g}$ in the outer atmosphere
(note that  log\,$P_{\rm g} \approx -0.5$ at the surface of our 
photospheric model shown in Fig.\,2). However, this result of course 
does not prove the validity of LTE assumption in molecular clouds. 
 Without a better method, however, we have assumed LTE to estimate
abundances of some other molecules of interest, for example TiO. For
$T = 2000$\,K and log\,$P_{\rm g} = -1.63$, TiO/CO is 4.09 x 10$^{-4}$. 
Then TiO column density is about 2 x 10$^{+16}$ cm$^{-2}$, which is 
not high enough to provide TiO features as strong  
as those noted by Ryde et al.\,(2003) for their cool photospheric model.  
For predicting molecular abundances in molecular clouds, detailed analysis 
of non-equilibrium processes are required, and 
we hope increasingly sophisticated approaches will be applied  
as more detail about the molecular clouds can be clarified.

In this paper, we concluded that the infrared spectra of cool luminous
stars cannot be understood as originating in the
photosphere alone, but should have hybrid nature due to
at least two components: a photosphere and a slightly cooler but 
not very cool molecular clouds or MOLsphere. We
noticed that ignorance of the hybrid nature of the infrared
spectra results in a large systematic error in abundance 
determinations based on molecular lines (Tsuji 2008).
Other authors (e.g. Decin et al. 2003) also noticed that
the observed infrared spectra of red giant stars, including Arcturus, 
are not necessarily matched very well by the predicted ones based on the 
most recent model photospheres. We suggest that such a difficulty  may 
also be related to the hybrid nature of the infrared  spectra, 
at least partly.  

By the way, we have found that the classical curve-of-growth method
 provides a simple means by which to recognize the hybrid
nature of the infrared spectra. The CG method can easily be applied to
any object only if EWs of some dozens of lines can be measured.
What is important is to measure as many lines as possible,
including the weak and the intermediate-strength lines.  
We hope that this simple CG method will be applied to other
objects (e.g. carbon stars) and/or to other spectral lines (e.g. TiO,
atomic lines) to clarify the nature of the hybrid spectra in more detail.  
Also, the simple CG method is effective when  spectral
lines consist of the contributions from different components
 as in the CO fundamental lines (Sect.\,4.1). 

Given that the molecular clouds have low  relative motions
and rather similar temperatures relative to the surrounding
atmosphere, they could be recognized only by subtle
spectroscopic signatures. For this very reason, the possible presence of 
such a molecular constituent has been overlooked for a long time. 
This is in  marked contrast to the case of a hot chromosphere, which
could be recognized in terms of the strong emission lines, and an 
expanding circumstellar envelope recognized by the blue-shifted 
zero-volt lines at an early time (Adams \& MacCormac 1935). 
Now do clouds exist in the atmosphere of Arcturus? It is 
certainly not easy to answer this question by spectroscopic observations 
alone.  But we found several arguments in favor of the presence of 
molecular clouds in Arcturus in this paper. Also, in cooler luminous 
stars in which CO lines show unusual behaviors as in Arcturus
(Sect.\,6.3), the presence of molecular clouds has been  shown more 
directly by observations with spatial interferometry (e.g. Mennesson 
et al. 2002; Perrin et al. 2004a). We hope further progress in 
sensitivity as well as in resolution in future stellar interferometry 
so that clouds in stellar atmospheres can be resolved.

The outer layers of red giant stars are quite complicated consisting
of many different components, and the definition of each component
is by no means clear. For example, the terms ``atmosphere'' and 
``photosphere'' are often confused. Probably, ``atmosphere'' may
involve all the observable outer layers of a star while ``photosphere''
represents the base of the atmosphere where continuum radiation and
most absorption lines are formed. We have no model
of an ``atmosphere'' for red giant star yet, even if we
have some models of the ``photosphere''. In the literature, such a
``model photosphere'' is often referred to as ``model atmosphere''
but, strictly speaking, this does not represent the real situation.

Thus, we have no model atmosphere of cool luminous stars yet and, 
moreover, the picture of 
stellar atmospheres might have been too simplified until the present. 
For example,  it may be an over-simplification to assume that the 
stellar atmospheres are completely free of clouds.  
Cloud formation is a rather common feature in celestial 
objects that we know in detail (e.g. our Earth and planets). The dust 
clouds are also common in brown dwarfs and 
dust cloud formation in ultracool dwarfs is now under intensive  
investigations by different approaches (e.g. Helling 
et al. 2008). Molecular cloud formation  in cool luminous stars 
is an important topic that should be pursued  in more detail,
since it can be one of the basic processes in cool stellar atmospheres,
and should have important effects on 
many problems from interpretation of the spectra to the origin of 
mass-loss. Molecular clouds in cool luminous stars
may  encompass  masering clouds observed in OH-IR stars, and thus
cloud formation should play a key role in understanding various
phenomena observed in all the family of cool luminous stars.     

A new problem that we have encountered is how the molecular clouds,  
referred to as MOLsphere for convenience in modeling, can be formed in 
the outer atmosphere of cool luminous stars. The MOLsphere differs 
from the so-called CO-mosphere if the CO-mosphere simply represents the 
cooler component of the bifurcated thermal structure, while the 
MOLsphere should consist of molecular clouds formed from additional 
gaseous material (Sect.\,6.1). Thus, the first problem is to 
explain how additional material can be supplied to 
the outer atmosphere. To push additional material into the upper 
atmosphere, magnetic forces due to MHD processes appeared to be
promising (Sect.\,6.4), and we hope that there will be further 
progress in this field. The second problem to be solved is how clouds 
form with the material supplied by dynamical processes. In an environment
where molecules and dust form easily (i.e. at rather low temperatures
in cool stellar atmospheres), cloud formation is possible,
for example,  by cooling instability induced by the
molecular formation itself (Sect.\,6.1). Thus it seems to be more 
natural to assume that clouds form in the atmosphere of cool stars 
than to assume that the atmosphere remains  free of any cloud. However, 
further detail of these processes should be worked out.      

To answer the  question ``Is Arcturus a well-understood K giant?'' 
(Verhoelst et al. 2005), we  unfortunately cannot answer positively 
at present. This question was addressed in connection with the near-IR 
interferometric observations, but spectroscopic observations are by no 
means more clearly understood yet. For observations such as the 
unexpected detection of H$_2$O pure rotation lines (Ryde et al. 2002) 
or unpredictable upturn of the flat part of the curves-of-growth for 
CO lines (Fig.\,5), we have no definite guidelines by which to 
interpret the observed data. The situation is more or less the same 
for other cool luminous stars, and relatively well-observed Arcturus 
will serve as invaluable reference for the interpretation and analysis 
of other cool luminous stars for which observations are more limited. 
Thus, we have sufficient reasons why we study Arcturus carefully.    

\begin{acknowledgements}
    I thank  Takeru K. Suzuki for helpful discussion on 
stellar mass-loss and related problems in red giant stars.
I am much indebted to an anonymous referee for invaluable 
suggestions and many helpful comments in improving the text.
My thanks are also due to  Kenneth H. Hinkle for making available 
the electronic version of the Arcturus Atlas at an early time. 
Data  analyses were in part carried out on common use data 
analysis computer system at the Astronomy Data Center (ADC) of 
NAOJ and this research has made use of the VizieR catalogue
access tool, CDS, Strasburg, France. This work was supported by 
Grant-in-Aid of JSPS for Scientific Research (C) nos.17540213 \& 
21540237.
\end{acknowledgements}

\end{document}